\documentclass[twocolumn,preprintnumbers,amsmath,amssymb,pra,superscriptaddress]{revtex4}
\usepackage{amsmath}
\usepackage{graphicx}
\usepackage{dcolumn}
\usepackage{bm}
\usepackage{units}
\usepackage{stmaryrd}
\usepackage[breaklinks=true,pdfborder={0 0 0}]{hyperref}
\usepackage[usenames]{color}
\usepackage{tabularx}

\usepackage{color}
\usepackage{tikz}
\usepackage{multirow}
\usepackage{amsfonts}
\usepackage{dsfont}
\usepackage{mathrsfs}
\usepackage{xcolor}
\usepackage{accents}
\newcommand{\ex}{\mathrm{e}}

\newcommand{\bd}{\bm{D}}
\newcommand{\br}{\bm{r}}
\newcommand{\bk}{\bm{k}}
\newcommand{\eh}{\hat{e}}
\newcommand{\be}{\bm{\mathcal{E}}}

\hypersetup{colorlinks=true,citecolor=red}
\newcommand{\ddd}[1]{{\color{blue}#1}}
\newcommand{\rrr}[1]{{\color{red}#1}}

\newlength{\LL} \LL 1\linewidth
\usepackage{soul,xcolor}
\setstcolor{red}
\begin{document}
\title{Comprehensive Quantum Framework for Describing Retarded and Non-Retarded Molecular Interactions in External Electric Fields}
\author{Mohammad~Reza~Karimpour}
\author{Dmitry~V.~Fedorov}
\author{Alexandre~Tkatchenko}
\email{alexandre.tkatchenko@uni.lu}
\affiliation{Department of Physics and Materials Science, University of Luxembourg, L-1511 Luxembourg City, Luxembourg}
\begin{abstract}
We employ various quantum-mechanical approaches for studying the impact of electric fields on both nonretarded and retarded noncovalent interactions between atoms or molecules. To this end, we apply perturbative and non-perturbative methods within the frameworks of quantum mechanics as well as quantum electrodynamics. In addition, to provide a transparent physical picture of the different types of resulting interactions, we employ a stochastic electrodynamic approach based on the zero-point fluctuating field. Atomic response properties are described via harmonic Drude oscillators --- an efficient model system that permits an analytical solution and has been convincingly shown to yield accurate results when modeling non-retarded intermolecular interactions. The obtained intermolecular energy contributions are classified as field-induced electrostatics, field-induced polarization, and dispersion interactions. The interplay between these three types of interactions enables the manipulation of molecular dimer conformations by applying transversal or longitudinal electric fields along the intermolecular axis.
Our framework based on the combination of four different theoretical approaches paves the way toward a systematic description and improved understanding of molecular interactions when molecules are subject to both external and vacuum fields.
\end{abstract}
\maketitle
\section{Introduction}
Molecular forces, stemming from noncovalent interactions between closed-shell atoms or molecules, govern various physical properties of different states of matter. These ubiquitous forces are widely present in systems throughout biology, chemistry, and physics, with an extension to practical fields such as nanotechnology and pharmaceuticals. For example, molecular forces play a major role in determining the structure, stability, and function for molecules and materials including proteins, nanostructures, molecular solids, and crystalline surfaces~\cite{Kaplan2006,Stone2013,Tkatchenko2015}. In general, atoms in a given molecule or material are subject to internal and external fields. 
These fields can be of static and/or dynamic origins and can arise either from neighboring atoms carrying effective partial charges within the same molecule or from external environments, such as cell membranes, ionic channels, liquids, among other possibilities. 
From an atom-wise perspective, the forces arising from the surrounding environment can be effectively modeled by external fields acting on an atom from all other components of the system. Hence, a complete theoretical description of interatomic interactions necessitates the modeling of arbitrary internal and external fields that atoms can experience.

The effect of static and dynamic external fields on noncovalent interactions has been studied using various theoretical frameworks for atomic and molecular systems. It was shown that by applying electromagnetic and thermal fields one can influence noncovalent interactions in several ways~\cite{Thirunamachandran1980, Rubio2021, Milonni1996, Marinescu1998, Sukhov2013, Brugger2015, Muruganathan2015, Kleshchonok2018, Fiscelli2020}. Random and nonuniform fields can affect the strength and distance scaling laws of the van der Waals (vdW) dispersion interaction or even change its sign~\cite{Thirunamachandran1980, Rubio2021, Milonni1996, Sukhov2013, Brugger2015, Sadhukhan2017}. Application of weak static (in)homogeneous fields to molecular dimers~\cite{Marinescu1998, Muruganathan2015, Kleshchonok2018} modifies the dispersion interactions in second and third orders of perturbation theory in the nonretarded regime, while the retarded regime was not addressed in these studies. The dispersion interactions under weak static fields in retarded regime were recently studied by Fiscelli \emph{et al}.~\cite{Fiscelli2020} using quantum electrodynamics (QED). They proposed a novel contribution to the retarded dispersion energy between two interacting two-level hydrogen atoms, scaling as $\propto R^{-4}$ with respect to the interatomic distance. Despite the relatively large number of studies on molecular interactions in electric fields, a comprehensive understanding of this topic is still missing and some results remain controversial. For instance, there is still an ongoing debate on the interpretation of vdW interactions in spatially-confined systems as having either an electrostatic or a quantum-mechanical origin~\cite{comment-to-Mainak, Mainak-reply, Stoehr2021}. 
The unusual $R^{-4}$ scaling of dispersion interactions in QED induced by an external field and recently presented in Ref.~\cite{Fiscelli2020} has also been debated as arising either from quantum effects or simply classical electrostatics~\cite{Comment_on_Fiscelli2020, Fiscelli_reply}. 
To resolve existing controversies and clarify discrepancies present in literature, in this \rrr{paper we} aim to develop a comprehensive framework for modeling and understanding molecular interactions in the presence of electric fields.
Our approach is firmly based on first principles of quantum mechanics (QM) and quantum electrodynamics and employs an exactly solvable model for the atomic polarization.

The QM theory of intermolecular forces is a well established field, with several seminal monographs covering this topic rather comprehensively~\cite{Kaplan2006,Stone2013}. Interactions between systems ranging from single atoms and small molecules to large macromolecules and nanostructures have been studied extensively in the nonretarded regime within QM framework~\cite{Szalewicz-SAPT-ChemRev-1994,Szalewicz-WIREs-2012,Tkatchenko2012,DiStasio2014,Reilly2015,Grimme-ChemRev2016,Hermann2017,Stoehr-CSR-2019}.
Typically, second-order intermolecular perturbation theory is employed to distinguish three types of noncovalent molecular interactions: electrostatic interactions between permanent multipoles, polarization (or induction) interactions between permanent and induced multipoles, and vdW dispersion interactions between induced fluctuating multipoles. When using higher orders of perturbation theory the situation is somewhat obscured because the distinction between different types of multipoles (permanent, induced, and fluctuating) becomes less obvious. The presence of electric fields, excitations, or specific boundary conditions introduces additional complications. Many of such field-induced phenomena are not yet part of textbook knowledge, even from the point of view of standard QM. For example, the qualitative change of vdW dispersion interactions induced by confinement of molecules in nanostructures or under inhomogeneous electric fields is a recent proposition~\cite{Sadhukhan2017,Kleshchonok2018,Stoehr2021}.

QED provides a well-established general framework to study the interaction between atoms (or small molecules) and the electromagnetic radiation field~\cite{Cohen-Tannoudji1997,Milonni1994,Craig1994,Greiner2008,Rubio2018}. There is a diversity of effects in QED that transcend standard QM interpretation and stem from the zero-point fluctuations of the electromagnetic radiation field. Such examples include vacuum polarization, self-energy terms, Lamb shift, and even particle creation and annihilation in strong fields~\cite{Milonni1994,Greiner2008}. QED has also been widely used for studying vdW dispersion and Casimir interactions between atoms and materials~\cite{Craig1994,Buhmann2013,Salam2009,Passante2018}. Owing to the relative complexity of the QED terms compared to their QM counterparts, one is often constrained to using effective models for the atomic response and its coupling to the quantum radiation field. Due to these reasons, the QED theory of molecular forces requires further development to reach the sophisticated level achieved by its QM analog. This fact is for example illustrated by the recent work of Fiscelli \textit{et al.}~\cite{Fiscelli2020}, which proposed the existence of a new QED dispersion energy term for two hydrogen atoms subjected to an external electric field.

With the aim to bridge the QM and QED treatments of molecular forces, in this work we develop a comprehensive framework and apply it to study the effect of a static electric field on noncovalent interactions between two atoms or molecules. To achieve a comprehensive understanding, we found it necessary to employ three different theories, given by molecular quantum mechanics, microscopic quantum electrodynamics, and stochastic electrodynamics. 
The three frameworks have been widely used in different communities in order to explore various aspects of intermolecular interactions. 
In contrast to QM and QED, the approach of stochastic electrodynamics~\cite{Marshall1963, Marshall1965, Boyer1975, Pena1978, Boyer1980, Pena1996, Pena2001, Pena2006, Nieuwenhuizen2019, Boyer2019}, as a classical stochastic approximation to QED, provides clear interpretations of different interaction terms for non-relativistic quantum-mechanical problems. It has been already shown that stochastic electrodynamics can successfully reproduce results of QED when studying vdW and Casimir-Polder interactions in absence of external fields for atomic and molecular systems~\cite{Boyer1969, Boyer1971, Boyer1972, Boyer1973, Marshall1992, Rueda1993}.

When studying the effect of external electric fields on interatomic interactions, the two-level ``hydrogen atom'' is often employed as a model system for atomic response. Unfortunately, this model system (two hydrogen-like atoms plus the external field) does not allow an analytical solution and this can lead to artifacts, especially when applying QED. To avoid this problem and to enrich our conceptual understanding of the effect of external fields on intermolecular interactions, in this work we employ the quantum Drude oscillator (QDO)~\cite{Wang2001,Sommerfeld2005,Jones2013, Sadhukhan2016} model for describing atomic and molecular responses in closed-shell systems. The usage of QDOs to accurately and efficiently model the response of valence electrons in atoms and molecules is a critical aspect because coupled QDOs enable analytical solutions, with and without electric field. In the next section, we briefly describe the well-known QDO model and its applications in studies of intermolecular interactions. 
Then, the problem of intermolecular interactions in a uniform static electric field is tackled by using four different approaches. 
Section III presents a detailed description of the exact diagonalization method, to derive an exact formula for the interaction energy, which then is approximated by compact expressions obtained using Taylor expansions. In addition to the leading contributions to the interaction energy discussed throughout the paper, (in Sections III and VI) we consider the full infinite-order series of interactions.
In Section IV, we show how the approximated results of Section III can be reproduced using perturbation theory in the framework of QM. To take into account the effect of retardation, we also employ QED and stochastic electrodynamics. The corresponding exhaustive derivation provides one with a guidance for practical uses of QED and stochastic electrodynamics as applied to coupled QDOs. Namely,
Section V contains a derivation of the interaction energy for both retarded and nonretarded regimes from a perturbative approach within the QED framework. To identify and interpret all dominant contributions to the interaction energy, in Section VI, we rationalize the results of the other three approaches by means of stochastic electrodynamics. To illustrate possible practical applications of the developed framework, in Section VII, we consider argon-argon and benzene-benzene dimers as two representative examples for atomic and molecular systems. Finally, we discuss the obtained results and make conclusions in Section VIII.

\section{Quantum Drude Oscillator Model for Atomic Polarization Response}
The harmonic oscillator~\cite{Bloch1997} is one of the exactly solvable systems in quantum mechanics. 
This fundamental model has been used in many branches of physics and chemistry including quantum field theory and quantum electrodynamics, quantum optics, statistical mechanics, solid-state physics, spectroscopy, and high-energy physics. The success of this model stems from the fact that the energy of physical systems near equilibrium can be well approximated by quadratic functions of variables representing displacements from the equilibrium state.
Especially, quantum harmonic oscillators are widely employed to describe the response of quantum-mechanical systems to weak external perturbations. 

As a representative of the class of models based on the quantum harmonic oscillator, the quantum Drude oscillator (QDO)~\cite{Wang2001,Sommerfeld2005,Jones2013, Sadhukhan2016} is a coarse-grained quantum-mechanical approach for describing the electronic response of valence electrons in atoms and molecules. 
Within the QDO model, each atom or molecule is represented by a Drude quasiparticle characterized by its mass $m$ and charge $(-q)$ bound to a nucleus of an opposite charge and an infinite mass through a harmonic potential with a characteristic frequency $\omega$. The three adjustable parameters of the QDO model can exactly reproduce a set of three atomic/molecular response properties. To properly capture the response of valence electrons, a reasonably accurate parameterization is~\cite{Jones2013}
\begin{align}
	\label{QDO-parameter}
	q=\sqrt{m\omega^2\alpha}\ , \ \ \ 
	m=\frac{5\hbar~ C_6}{\omega~ C_8} \ , \ \ \ 
	\omega=\frac{4C_6}{3\hbar\alpha^2} \ ,
\end{align} 
obtained by reproducing the dipole polarizability $\alpha$ as well as the $C_6$ and $C_8$ dispersion coefficients of homospecies dimers taken from experimental or calculated \textit{ab initio} reference data for atoms or molecules. 
When adjusting the QDO parameters to accurate reference data, this coarse-grained model constitutes a simple yet efficient tool to describe response properties and non-covalent interactions of atoms, small and large (bio)molecules, solids, nanostructures and hybrid organic/inorganic interfaces~\cite{Wang2001,Sommerfeld2005,Jones2013,Tkatchenko2012,Reilly2015,DiStasio2014,Gobre2016,Sadhukhan2016,Hermann2017,Fedorov2018,Tkatchenko2020,Vaccarelli2021}.
Specifically, the QDO model can quantitatively -- within a few percent compared to explicit treatment of electrons -- describe polarization and dispersion interactions~\cite{Jones2013,Sadhukhan2016,Hermann2017} as well as accurately capture electron density redistribution induced by these interactions~\cite{Hermann-NatureComm}. 
In addition, QDOs have been shown to provide a robust tool to describe vdW interactions under the influence of external charges as well as spatial confinement~\cite{Sadhukhan2017,Kleshchonok2018,Stoehr2021}. 
Finally, even though the QDO model describes distinguishable Drude particles bound to their own nuclei, it is possible to generalize this model to quantum bosonic statistics. 
Introducing Pauli-like exchange interactions to the QDO model allowed to derive a generalized quantum-mechanical relation between atomic polarizabilities and van-der-Waals radii, demonstrating its validity for many atoms in the periodic table~\cite{Fedorov2018,Tkatchenko2020,Vaccarelli2021}.

The present work benefits from the quadratic form of the QDO Hamiltonian which allows diagonalization of the Hamiltonian of a system of interacting QDOs with or without an external field being applied. 
Using the dipole approximation for the atom-atom and atom-field couplings, such an exact diagonalization procedure yields a new system of decoupled QDOs, whose ground state contains all the dipolar interaction terms. The importance of this self-consistent solution grows with size and complexity of the system containing many interacting species~\cite{Tkatchenko2012, DiStasio2014, Reilly2015, Gobre2016}. 
On the other hand, the complete set of eigenstates of a QDO in a uniform electric field enables expanding perturbed states of the coupled QDO--field system under the influence of linear perturbations, {\it e.g.} describing interactions with nearby QDOs as well as with macroscopic bodies and boundary conditions. In turn, such an expansion allows one to study retarded and nonretarded field-mediated intermolecular interactions by means of the perturbation theory within QED. 

Despite all the compelling analytical and computational features offered by the QDO model and its extensive applications in QM theory of intermolecular interactions, this model has not been widely used in molecular QED. A certain connection has been established in the work of Ciccarello {\it et al.}~\cite{Passante2005} who have shown, by means of a nonperturbative approach, that describing two identical atoms by charged harmonic oscillators can reproduce the well-known Casimir-Polder energy for the retarded dispersion interaction. In the present work, we substantially advance the use of QDOs within the QED framework for studying interactions of atomic and molecular systems with electromagnetic fields and/or other atoms and molecules.

The exact results for two dipole-coupled QDOs in static electric fields, which we present in the next section, can be straightforwardly generalized to an arbitrary number of interacting species. This feature of the QDO model allows one to easily extend the existing many-body approaches for description of vdW interactions~\cite{Tkatchenko2012,DiStasio2014} to include external fields. Moreover, such an approach enables numerically exact descriptions of the effect of intra-molecular fields on molecular polarizabilities: considering atomic charge redistribution in a molecule due to local electric fields caused by interactions with other atoms, one can accurately obtain molecular polarizabilities based on hybridized (atom-in-a-molecule) polarizabilities of constituting atoms.

\section{Molecular quantum mechanics: Exact diagonalization}
In this section we present a nonperturbative approach for describing the interaction between two species (atoms or molecules) in the presence of a uniform static electric field. We make use of the exact solution of the QDO model in both cases where the QDO is either coupled via its electric dipole moment to another QDO or subject to an external static electric field. Using a two-step normal-mode transformation, this allows us to diagonalize the total Hamiltonian for a system of two interacting QDOs which are initially coupled to an external field.

\begin{figure}[t]
\includegraphics[width=0.85\linewidth]{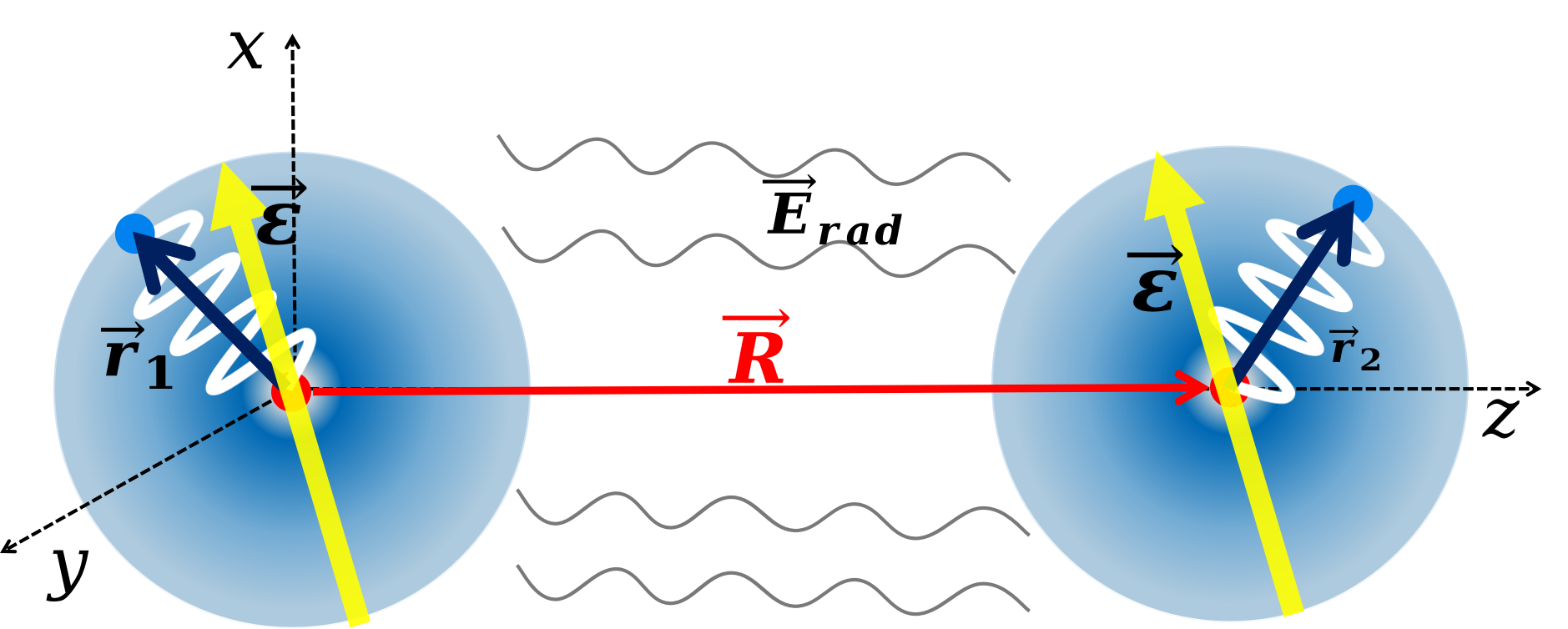}
\caption{Two interacting atoms or molecules represented as quantum Drude oscillators (QDOs), separated by a distance $R=|\vec{\bm{R}}|$ along the $z$ axis, under the influence of both, the fluctuating vacuum electromagnetic field $\vec{\bm{E}}_{rad}$ and an applied uniform static electric field $\vec{\bm{\mathcal{E}}} = (\mathcal{E}_x,\mathcal{E}_y,\mathcal{E}_z)$.}
\label{fig:QDOs}
\end{figure}

In the nonretarded regime, when the interspecies distance $R$ is much smaller than the characteristic wavelength $\lambda_e$ of electron transitions to excited states, $\lambda_e\!\gg\!R$, the interaction reduces to the instantaneous Coulomb coupling. Thus the Hamiltonian of a system of two interacting QDOs reads
\begin{align}
\label{nret-Hamiltonian-1}
H = \sum_{i=1,2}\left[-\frac{\hbar^2}{2m_i}\bm{\nabla}_{\bm{r}_i}^2+\frac{1}{2} m_i\omega_i^2 \bm{r}_i^2\right]+V(\bm{r}_1,\bm{r}_2)\ .
\end{align}
Here, $m_i$ and $\omega_i$ are masses and characteristic frequencies of the two Drude particles~\cite{Jones2013}, respectively. If the interacting QDOs are located along the $z$ axis and separated by the distance $R$ (see Fig.~\ref{fig:QDOs}), then the coupling Coulomb potential in its dipole approximation is
\begin{align}
\label{nret-V}
\!\!\!\!\!
V(\bm{r}_1,\bm{r}_2) \approx V_{\rm dip} (\bm{r}_1,\bm{r}_2) = \frac{q_1 q_2}{(4\pi\epsilon_0) R^3} \left(\bm{r}_1\cdot\bm{r}_2 - 3z_1 z_2 \right) ,\!\!\!\!
\end{align}
where $-q_i$ is the charge of $i$th Drude particle bound to its nucleus with the charge $q_i\,$. Then, the $x$-dependent part of the Hamiltonian in Eq.~\eqref{nret-Hamiltonian-1} is given by
\begin{align}
\label{nret-Hamiltonian-x1}
\!\!\!
H_x=\!\!\!\!
\sum_{i=1,2}\!\left[
-\frac{\hbar^2}{2m_i}\frac{d^2}{d x_i^2}
+\frac{1}{2}m_i \omega_i^2 x_i^2
\right]
+ \frac{q_1 q_2}{(4\pi\epsilon_0) R^3} x_1 x_2\ .\!\!
\end{align}
Introducing new coordinates $x'_1 = \sqrt{m_1} x_1$ and $x'_2=\sqrt{m_2} x_2\,$, Eq.~\eqref{nret-Hamiltonian-x1} transforms to
\begin{align}
\label{nret-Hamiltonian-x2}
H_x = 
\sum_{i=1,2}\!\left[
-\frac{\hbar^2}{2}\frac{d^2}{d {x'_i}^2}
+ a_i {x'_i}^2
\right]
+ \gamma_x\, x'_1 x'_2\ ,
\end{align}
with $\gamma_x=\frac{q_1 q_2}{(4\pi\epsilon_0)\sqrt{m_1 m_2} R^3}$ and $a_i =\omega_i^2/2$. To diagonalize this Hamiltonian, we rewrite the potential energy in a matrix form
\begin{align}
a_1 {x'_1}^2 + a_2 {x'_2}^2 + \gamma_x x'_1 x'_2 
= \left( x'_1\ \ x'_2 \right) \hat{\bm{M}}
\left(
\begin{matrix}
x'_1\\
x'_2
\end{matrix}
\right)\ ,
\end{align}
where $\hat{M}_{ii} = a_i$ and $\hat{M}_{12} = \hat{M}_{21} = \gamma_x/2$.
The eigenvalues and orthonormal eigenvectors of matrix $\hat{\bm{M}}$ are
\begin{align}
\lambda_\pm = \frac{1}{2}\left[(a_2 + a_1) \pm \sqrt{D_x}\right]
\end{align}
and
\begin{align}
c_{\pm}= \frac{1}{A_\pm} 
\left(\begin{matrix}
\gamma_x \\
(a_2 - a_1)\pm\sqrt{D_x}
\end{matrix}\right) ,
\end{align}
respectively. Here, we have employed the notations
\begin{align}
\begin{array}{ll}
\hspace{-0.15cm}
D_x = (a_2 - a_1)^2 + \gamma_x^2\, , \, 
A_\pm = \sqrt{\gamma_x^2 + [(a_2 - a_1) \pm \sqrt{D_x}]^2}\, .
\end{array}
\end{align}
Introducing the normal-mode coordinates
\begin{align}
\label{new_coordinates}
x_\pm &=\frac{1}{A_\pm}\left(\gamma_x x'_1 + [(a_2 - a_1) \pm \sqrt{D_x}] x'_2\right)
\end{align}
and making use of the coordinate transformation
\begin{align}
\label{transforme}
&x'_1=\frac{\gamma_x}{A_+}x_++\frac{\gamma_x}{A_-}x_- \ ,
\\ \nonumber
&x'_2=\frac{(a_2 - a_1)+\sqrt{D_x}}{A_+}x_++\frac{(a_2 - a_1) - \sqrt{D_x}}{A_-}x_-\ ,
\end{align}
one can diagonalize the Hamiltionian of Eq.~\eqref{nret-Hamiltonian-x2} by expressing it in terms of the normal-mode coordinates $x_\pm$ and corresponding frequencies $\omega_\pm = [(a_1 + a_2) \pm \sqrt{D_x}]^{\nicefrac 12}$ as
\begin{align}
\label{nret-Hamiltonian-x3}
H_x = \sum_{i=\pm}-\frac{\hbar^2}{2}\frac{d^2}{dx_i^2}+\frac{1}{2}
\omega_i^2x_i^2\ .
\end{align}
Equation~\eqref{nret-Hamiltonian-x3} is the Hamiltonian of a system of two uncoupled QDOs with frequencies $\omega_\pm$ and unit masses.

Now we apply an external uniform static electric field, $\bm{\mathcal{E}} = (\mathcal{E}_x,\mathcal{E}_y,\mathcal{E}_z)$, to this system by introducing the field--QDOs coupling Hamiltonian in the dipole approximation 
\begin{equation}
H_{\rm f} = - (q_1 \bm{r}_1 + q_2 \bm{r}_2)\! \cdot\! \bm{\mathcal{E}}\ .
\end{equation}
For its $x$-dependent part, we have
\begin{align}
\label{field-coupling-x-1}
\!\!\!H_{{\rm f}, x} = - (q_1 x_1 + q_2 x_2) \mathcal{E}_x
= - \left(\frac{q_1 x_1^\prime}{\sqrt{m_1}} + \frac{q_2 x_2^\prime}{\sqrt{m_2}} \right) \mathcal{E}_x\ .\!
\end{align}
On making use of the transformations of Eq.~\eqref{transforme} the above Hamiltonian becomes
\begin{align}
\hspace{-0.25cm}
H_{{\rm f}, x} = - \sum_{i=\pm} f_i x_i\mathcal{E}_x\ ,
\end{align}
where the prefactors $f_+$ and $f_-$ are defined as
\begin{align}
f_\pm = \frac{1}{A_\pm}\left(\frac{q_1}{\sqrt{m_1}}\gamma_x + \frac{q_2}{\sqrt{m_2}} [(a_2 - a_1)\pm \sqrt{D_x}]\right).
\end{align}
Therefore, the $x$-component of the total Hamiltonian, 
$H' =H + H_{\rm f}\,$, for the two interacting QDOs in the presence of an external uniform static electric field reads
\begin{align}
\label{nret-Hamiltonian-x4}
H'_x =\sum_{i=\pm} \left[ -\frac{\hbar^2}{2}\frac{d^2}{d x_i^2}
+\frac{1}{2} \omega_i^2 x_i^2 -\mathcal{E}_x f_i x_i\right]
\ .
\end{align}
Completing the squares for $x_\pm$ yields the quadratic form
\begin{align}
\label{nret-Hamiltonian-x5}
H'_x\!=\! \sum_{i=\pm}\! \left[ -\frac{\hbar^2}{2} \frac{d^2}{d x_i^2}+ \frac{\omega_i^2}{2} \left(x_i-\frac{f_i\mathcal{E}_x}{\omega_i^2}\right)^2-\frac{f_i^2 \mathcal{E}_x^2}{2\,\omega_i^2}\right]\ ,\!\!
\end{align}
that can be considered as the Hamiltonian of two non-interacting 
one-dimensional (1D) oscillators with the characteristic
frequencies $\omega_\pm$ and shifted centers of oscillations
by the field-dependent factors $f_\pm\mathcal{E}_x/\omega_\pm^2\,$. Therefore, the ground-state energy corresponding to the Hamiltonian of Eq.~\eqref{nret-Hamiltonian-x5}
can be easily obtained as
\begin{align}
\label{nret-energy-in-f-x}
\mathscr{E}_{x}=\sum\limits_{i=\pm} \left[\frac{\hbar\omega_i}{2}
-\frac{f_i^2 \mathcal{E}_x^2}{2\,\omega_i^2}\right]\ .
\end{align}
On the other hand, the ground-state energy of the two non-interacting QDOs in the external field is the sum
\begin{align}
\label{nret-energy-in-f-x-isolated}
\mathscr{E}_{x}^{\rm (ni)} = 
\sum\limits_{i=1,2} \left[\frac{\hbar\omega_i}{2}
-\frac{\alpha_i\mathcal{E}_x^2}{2}\right]\ ,
\end{align}
with $\alpha_i = q_i^2 / m_i \omega_i^2$ as the isotropic static dipole polarizability of the $i$th isolated QDO. Comparing Eqs.~\eqref{nret-energy-in-f-x} and \eqref{nret-energy-in-f-x-isolated}, we see that $f_\pm$ play the role of the ratio $q/\sqrt{m}$ renormalized for the use of the collective (normal-mode) coordinates $x_\pm$ introduced in Eq.~\eqref{new_coordinates}.

In our next step we derive the interaction energy of the two 1D oscillators under the external electric field as the difference between the total energy of the coupled QDOs and the sum of the total energies of two isolated QDOs in the same field. Based on Eqs.~\eqref{nret-energy-in-f-x} and \eqref{nret-energy-in-f-x-isolated},  this interaction energy can be obtained as 
\begin{widetext}
\begin{align}
\label{nret-delta-energy-x}
\Delta & \mathscr{E}_x =
\mathscr{E}_x - \mathscr{E}_x^{\rm (ni)} =
\frac{\mathcal{E}_x^2}{1-{\alpha_1 \alpha_2}/([4\pi\epsilon_0]^2 R^6)}
\left( \frac{\alpha_1 \alpha_2}{[4\pi\epsilon_0] R^3}
-\frac{\alpha_1^2 \alpha_2}{2 [4\pi\epsilon_0]^2 R^6} - \frac{\alpha_1 \alpha_2^2}{2 [4\pi\epsilon_0]^2 R^6} \right)
-\frac{\hbar}{2} (\omega_1+ \omega_2) +\frac{\hbar \sqrt{2}}{4}\times \\ \nonumber \times&
\left\{ \sqrt{(\omega_1^2+\omega_2^2) + (\omega_2^2 - \omega_1^2)
\sqrt{1 + \frac{4\, \alpha_1 \alpha_2\, \omega_1^2\, \omega_2^2}{[4\pi\epsilon_0]^2  \left(\omega_2^2-\omega_1^2\right)^2 R^6}}} +
\sqrt{(\omega_1^2+\omega_2^2) - (\omega_2^2 - \omega_1^2)\sqrt{1 + \frac{4\, \alpha_1 \alpha_2\, \omega_1^2\, \omega_2^2}{[4\pi\epsilon_0]^2  \left(\omega_2^2-\omega_1^2\right)^2 R^6}}}  \right\}\ . 
\end{align}
Due to the symmetry of the considered system, $\Delta\mathscr{E}_y$ can be  obtained in the same way as $\Delta\mathscr{E}_x$ just by replacing the subscript $x$ with $y$. To derive $\Delta\mathscr{E}_z\,$, one needs to diagonalize the  $z$-dependent part of the total Hamiltonian, $H_z$, using similar transformations as given in Eq.~\eqref{transforme} but with replacing $x$ by $z$ and $\gamma_x$ by $\gamma_z=-2\gamma_x$. For this case, we obtain 
$z_\pm =\left(\gamma_z z'_1 + [(a_2 - a_1) \pm \sqrt{D_z}] z'_2\right)/\sqrt{\gamma_z^2 + [(a_2 - a_1) \pm \sqrt{D_z}]^2}$ and $\omega_\pm = [(a_1 + a_2) \mp \sqrt{D_z}]^{\nicefrac 12}$, where $D_z = (a_2 - a_1)^2 + \gamma_z^2$. Altogether, this leads to
\begin{align}
\label{nret-delta-energy-z}
\Delta & \mathscr{E}_z =
\mathscr{E}_z - \mathscr{E}_z^{\rm (ni)} =
- \frac{2\,\mathcal{E}_z^2}{1-{4\,\alpha_1 \alpha_2}/([4\pi\epsilon_0]^2 R^6)}
\left( \frac{\alpha_1 \alpha_2}{[4\pi\epsilon_0] R^3}
+\frac{\alpha_1^2 \alpha_2}{[4\pi\epsilon_0]^2 R^6} + \frac{\alpha_1 \alpha_2^2}{[4\pi\epsilon_0]^2 R^6} \right)
-\frac{\hbar}{2} (\omega_1+ \omega_2) +\frac{\hbar \sqrt{2}}{4}\times \\ \nonumber \times&
\left\{ \sqrt{(\omega_1^2+\omega_2^2) + (\omega_2^2 - \omega_1^2)
\sqrt{1 + \frac{16\, \alpha_1 \alpha_2\, \omega_1^2\, \omega_2^2}{[4\pi\epsilon_0]^2  \left(\omega_2^2-\omega_1^2\right)^2 R^6}}} +
\sqrt{(\omega_1^2+\omega_2^2) - (\omega_2^2 - \omega_1^2)\sqrt{1 + \frac{16\, \alpha_1 \alpha_2\, \omega_1^2\, \omega_2^2}{[4\pi\epsilon_0]^2  \left(\omega_2^2-\omega_1^2\right)^2 R^6}}} \right\}\ .
\end{align}
Despite their intricate form, formulas given by 
Eqs.~\eqref{nret-delta-energy-x} and \eqref{nret-delta-energy-z} allow us to clearly distinguish between electrostatic, polarization, and dispersion interactions. Indeed, the dispersion interaction energy results from the difference between the first terms of Eqs.~\eqref{nret-energy-in-f-x} and \eqref{nret-energy-in-f-x-isolated}.
Therefore, its contribution is proportional to the (reduced) Planck constant. Although the exact expressions for the dispersion energy in Eqs.~\eqref{nret-delta-energy-x} and \eqref{nret-delta-energy-z} do not allow to explicitly eliminate the distance-independent terms corresponding to the self-energies, all such terms cancel each other when we perform a Taylor expansion to obtain Eqs.~\eqref{nret-delta-energy-x-leading} and \eqref{nret-delta-energy-z-leading}.
In contrast to the dispersion energy depending on the characteristic frequencies of the interacting species,  the electrostatic and polarization  contributions to $\Delta  \mathscr{E}_x$, $\Delta  \mathscr{E}_y$, and $\Delta \mathscr{E}_z$ are fully determined by the two dipole polarizabilities, $\alpha_1$ and $\alpha_2$. The corresponding three terms in the large parentheses within the first line of Eqs.~\eqref{nret-delta-energy-x} and \eqref{nret-delta-energy-z} are the field-induced dipole-dipole electrostatic energy and two (symmetric) contributions to the field-induced polarization energy. 
The fraction in front of these parentheses encodes a mutual self-consistent polarization of two polarizable species under the external static field. 
By performing a Taylor expansion for this fraction, as we do below, one obtains an infinite series. 
As shown in Section VI, this series can be interpreted as a sum of interaction energies of an infinite number of dipole moments induced at the two QDOs, starting with the two initial dipoles, $\bm{\mu}_1 = \alpha_1 \bm{\mathcal{E}}$ and $\bm{\mu}_2 = \alpha_2 \bm{\mathcal{E}}$, induced by the applied electric field. 
The physical mechanism of the electrostatic/polarization infinite series is similar to the one known for the dispersion interaction~\cite{TAD-JCP2013}. 
The only difference is that the dispersion/polarization coupling originates from the fluctuating electric dipoles instead of the static dipoles relevant for the electrostatic/polarization series.

Taking into account that $\Delta \mathscr{E}_y = \Delta \mathscr{E}_x$, Eqs.~\eqref{nret-delta-energy-x} and \eqref{nret-delta-energy-z} provide one with the complete description of the total interaction energy between two QDOs in the presence of the external field. 
Due to the use of the QDO model, both the electrostatic/polarization and dispersion/polarization contributions are given by analytical formulas. 
However, to obtain more transparent expressions, we perform Taylor expansions of the first fraction within the first line as well as
the square roots within the second line in Eqs.~\eqref{nret-delta-energy-x} and \eqref{nret-delta-energy-z}.
These series expansions, performed below with respect to small terms proportional to $\alpha_1\alpha_2/(4\pi\epsilon_0)^2 R^6$, are related to the following physical picture. The employed dipole approximation for the Coulomb potential implies that the separation distance is much larger than the electronic clouds of two interacting species modeled by the QDOs. The effective radius of these clouds can be roughly described by $[\alpha/(4\pi\epsilon_0)]^{\nicefrac 13}$. This gives us the small parameter for the expansions, $\alpha_1\alpha_2/(4\pi\epsilon_0)^2 R^6 \ll 1$, and we obtain
\begin{align}
\label{nret-delta-energy-x-leading}
\Delta \mathscr{E}_{x} = & 
\frac{\alpha_1 \alpha_2\, \mathcal{E}_{x}^2}{[4\pi\epsilon_0] R^3} 
- \frac{\alpha_1 \alpha_2 (\alpha_1 + \alpha_2) \mathcal{E}_{x}^2}{2 [4\pi\epsilon_0]^2 R^6}
+\frac{\alpha_1^2 \alpha_2^2\, \mathcal{E}_{x}^2}{[4\pi\epsilon_0]^3 R^9} 
-\frac{\alpha_1^2 \alpha_2^2 (\alpha_1 + \alpha_2) \mathcal{E}_{x}^2}{2 [4\pi\epsilon_0]^4 R^{12}} 
\nonumber \\
-& \frac{\alpha_1 \alpha_2\, \hbar\, \omega_1\, \omega_2}{4 [4\pi\epsilon_0]^2 (\omega_1+\omega_2) R^6} 
-\frac{\alpha_1^2 \alpha_2^2\, \hbar\, \omega_1\, \omega_2\,(\omega_1^2+3\omega_1\omega_2+\omega_2^2)}{16 [4\pi\epsilon_0]^4 (\omega_1+\omega_2)^3 R^{12}}
+ O\left(1/R^{15}\right)\ ,
\end{align}
and
\begin{align}
\label{nret-delta-energy-z-leading}
\Delta \mathscr{E}_z = & 
- \frac{2\, \alpha_1 \alpha_2\, \mathcal{E}_z^2}{[4\pi\epsilon_0] R^3}
- \frac{2\, \alpha_1 \alpha_2 (\alpha_1 + \alpha_2) \mathcal{E}_z^2}{[4\pi\epsilon_0]^2 R^6}
- \frac{8\, \alpha_1^2 \alpha_2^2\, \mathcal{E}_z^2}{[4\pi\epsilon_0]^3 R^9} 
- \frac{8\, \alpha_1^2 \alpha_2^2 (\alpha_1 + \alpha_2) \mathcal{E}_z^2}{[4\pi\epsilon_0]^4 R^{12}} 
\nonumber \\
- & \frac{\alpha_1 \alpha_2\, \hbar\, \omega_1\, \omega_2}{ [4\pi\epsilon_0]^2 (\omega_1+\omega_2) R^6} 
-\frac{\alpha_1^2 \alpha_2^2\, \hbar\, \omega_1\, \omega_2\,(\omega_1^2+3\omega_1\omega_2+\omega_2^2)}{ [4\pi\epsilon_0]^4 (\omega_1+\omega_2)^3 R^{12}}
+ O\left(1/R^{15}\right)\ ,
\end{align}
where we have explicitly written the terms up to $R^{-12}$. 
The frequency-dependent terms, within the second line of Eqs.~\eqref{nret-delta-energy-x-leading} and \eqref{nret-delta-energy-z-leading}, correspond to the expansion of the dispersion energy which is already well known~\cite{TAD-JCP2013}. 
The first term in these equations describes the electrostatic interaction of two dipoles which are initially induced by the applied static field. 
Each of these initial field-induced dipoles produces its own electric field experienced by another QDO. 
In its turn, this additional electric field from one field-induced dipole induces a concomitant dipole on the other QDO.
The energy of such dipoles in the fields inducing them is given by the second term of Eqs.~\eqref{nret-delta-energy-x-leading} and \eqref{nret-delta-energy-z-leading} describing the field-induced polarization interaction. 
The interpretation of higher-order terms in this infinite series becomes more transparent within a semiclassical approach, as we show in section VI by using stochastic electrodynamics.

The higher-order electrostatic/polarization terms shown in Eqs.~\eqref{nret-delta-energy-x-leading} and \eqref{nret-delta-energy-z-leading} will have an important role for many-body interactions in large molecular systems~\cite{Gobre2013}. 
However, for the two-species system considered here, we are mainly interested in leading contributions up to $\propto R^{-6}$. 
Therefore, for now we neglect terms $\propto R^{-9}$ and higher-order contributions. 
This yields the total interaction energy between two QDOs under a static field as 
\begin{align}
\label{nret-total-int-dissimilar}
\Delta \mathscr{E} = 
\frac{\alpha_1 \alpha_2 (\mathcal{E}_x^2+\mathcal{E}_y^2-2\mathcal{E}_z^2)}{[4\pi \epsilon_0] R^3} - \frac{\alpha_1 \alpha_2(\alpha_1+\alpha_2) (\mathcal{E}_x^2+\mathcal{E}_y^2+4\mathcal{E}_z^2)}{2 [4\pi\epsilon_0]^2 R^6}
- \frac{3\, \alpha_1 \alpha_2 \, \hbar \, \omega_1 \, \omega_2}{2 [4\pi\epsilon_0]^2 (\omega_1+\omega_2) R^6}
\ .
\end{align}
\end{widetext}
An extension of this result to the case of anisotropic molecules (see Appendix A) is straightforward following the derivation presented in the current section. 
The last term in Eq.~\eqref{nret-total-int-dissimilar}
corresponds to the well-known nonretarded vdW dispersion interaction, which is not affected by the uniform static field. 
The first and second terms of Eq.~\eqref{nret-total-int-dissimilar} are field-induced electrostatic and polarization interaction energies, respectively. According to Eq.~\eqref{nret-total-int-dissimilar}, the field-induced electrostatic interaction can be attractive or repulsive depending on the orientation of the external static electric field with respect to the interspecies distance. By contrast, the field-induced polarization and dispersion interactions are always attractive. 
The different contributions to the interaction energy as well as the interplay between them will be discussed in more detail after the consideration of the case of large interspecies distances in comparison to the characteristic wavelengths of electron transitions to excited states, $R\gg \lambda_e $. 

As shown in the next section, the interaction energy given by Eq.~\eqref{nret-total-int-dissimilar}, which is an approximation to the exact result of Eqs.~\eqref{nret-delta-energy-x} and \eqref{nret-delta-energy-z}, can be derived from the Rayleigh-Schr\"odinger perturbation theory. 
However, before moving to this alternative approach, let us point out a noteworthy aspect of the considered exact diagonalization method which is not present for other approaches employed in our work. 
The opportunity to diagonalize the Hamiltonian of Eq.~\eqref{nret-Hamiltonian-x4}, as achieved in Eq.~\eqref{nret-Hamiltonian-x5}, implies that using the QDO model one can also easily capture the effect of intramolecular fields acting on atoms in a molecule. 
Indeed, covalent interactions cause charge transfer between atoms, which leads to a distribution of local centers of positive and negative charge over the molecular system. 
The ensuing electric fields can be regarded as local external fields acting on atoms (see Appendices B and C for dissimilar local fields applied to anisotropic QDOs). Using our exact diagonalization method, one can take into account the effect of such fields via spatial shifts of the center of QDOs describing atoms together with Stark shifts in atomic energies, 
in order to properly describe molecular polarizabilities. Such a self-consistent procedure applied to an arbitrary number of QDOs in an inhomogeneous electric field should allow to develop quantum-mechanical force-fields that can efficiently describe all types of intermolecular interactions in atomic and molecular systems.

\section{Perturbation theory in quantum mechanics}
Perturbation theory is a powerful and insightful tool in quantum mechanics and quantum electrodynamics, in particular for the calculation of molecular interaction energies. Within this approach, the quantum states of a system of interacting atoms or molecules can be expanded in the basis of non-interacting states and the interaction potentials are obtained as corrections to the total energy of non-interacting species. 
The application of perturbation theory requires the states of the unperturbed system to form a complete basis set.
Since a QDO in a static electric field is an exactly solvable
quantum-mechanical problem, we apply the Rayleigh-Schr\"odinger perturbation theory considering a system of two non-interacting QDOs in an external field as the unperturbed system. Then, the Coulomb interaction between the two QDOs plays the role of a perturbing potential. 
Assuming that the two QDOs are placed along the $z$ axis and separated by a distance $R$, the total Hamiltonian can be written as $H=H_0+V_{\rm int}\,$, 
where $V_{\rm int}$ in the dipole approximation is given by Eq.~\eqref{nret-V} and the Hamiltonian of the unperturbed system reads
\begin{align}
\label{H_0_21}\!\!\!\!\!
H_0=\sum_{i=1}^2 H^{(0)}_i
=\sum_{i=1}^2 \left(\frac{\bm{p}_i^2}{2m_i}+\frac{1}{2}m_i\omega_i^2\br_i^2-q_i\br_i\cdot\be \right) .\!\!\!
\end{align}
To obtain the eigenstates and eigenvalues of $H_i^{(0)}$, we diagonalize it by means of the auxilary transformation 
$\bm{r}=\bm{r}'+{q \be}/{m \omega ^2}$ giving us the auxilary Hamiltonian 
\begin{equation}
\label{auxilary_H}
{H'}_{\!\!i}^{(0)}=\frac{{\bm{p}'}_{\!\!i}^2}{2m_i}+\frac{1}{2}m_i\omega_i^2{\bm{r}'}_{\!\!i}^2 - \frac{1}{2}\alpha\mathcal{E}^2 \ ,
\end{equation}
where we actually have ${\bm{p}_i}' = {\bm{p}_i}$\,. This Hamiltonian corresponds to a quantum harmonic oscillator with energy levels shifted by $-\alpha\mathcal{E}^2/2$, due to the Stark effect, and the well-known wavefunctions~\cite{Atkins_Friedman_book}, which we denote here by $\phi_{\bm{n}}(\bm{r}')$. 
Thus, one can straightforwardly see that the eigenstates and eigenvalues of each unperturbed Hamiltonian $H_i^{(0)}$ in the actual coordinates $\bm{r}_i$ are given by
\begin{subequations}
\label{QDO-in-exF}
\begin{align}
\label{QDO-in-exF1}	
&\psi_{\{n_x,n_y,n_z\}}(\br_i) 
=\phi_{\{n_x,n_y,n_z\}}\left(\br_i-\frac{q \be}{m  \omega ^2}\right)\ , \\
\label{QDO-in-exF2}
&E_{\{n_x,n_y,n_z\}}=\hbar\omega\left(n_x+n_y+n_z+\frac{3}{2}\right)-\frac{1}{2}\alpha\mathcal{E}^2\ .
\end{align}
\end{subequations}
Throughout our discussion below, we refer to the two QDOs with Hamiltonians $H_1^{(0)}$ and $H_2^{(0)}$ and the shifted wavefunctions
and energy eigenvalues as \textit{unperturbed QDOs}.
Using the wavefunctions of Eq.~\eqref{QDO-in-exF}, one can calculate matrix elements of the electric dipole operator, $\bm{\mu} = q\,\bm{r}\,$.
For the $x$ component of the dipole moment, we have
\begin{align}
\label{xmatel}
\langle j|\mu_x|i\rangle
=&q\langle j^{(0)}|x|i^{(0)}\rangle+\langle j^{(0)}|\frac{q^2\mathcal{E}_x}{m \omega^2}|i^{(0)}\rangle \\ \nonumber
=&q\sqrt{\tfrac{\hbar}{2m\omega}}\left[\textstyle{\sqrt{j}}\,\delta_{j,i+1}+\textstyle{\sqrt{j+1}}\,\delta_{j,i-1}\right] + \alpha\,\mathcal{E}_x\,\delta_{i,j}\ ,
\end{align}
where $\langle x|i\rangle=\psi_i(x)$ and $\langle x|i^{(0)}\rangle=\phi_i(x)$, with $\psi$ and $\phi$ introduced in Eq.~\eqref{QDO-in-exF}. The $y$ and $z$ components of the dipole moment can be obtained similarly.

Having the eigenstates and eigenvalues of $H_i^{(0)}$, the wavefunctions of $H_0$ in Eq.~\eqref{H_0_21} can be written as product states
$\Psi(\bm{r}_1,\bm{r}_2)=\psi(\bm{r}_1)\psi(\bm{r}_2)$. 
In what follows, we calculate the energy shifts due to the Coulomb coupling between the two QDOs up to the second-order correction using the matrix elements of the atomic dipole moments given by Eq.~\eqref{xmatel}.

From the first-order perturbation, the energy shift is
\begin{align}
\label{ES-pert-mqm}
&\Delta \mathscr{E}^{(1)}={}_2\langle 0,0,0|~{}_1\langle 0,0,0|~V_{int}~|0,0,0\rangle_1~|0,0,0\rangle_2
\nonumber\\
&=\frac{q_1^2 q_2^2\left(\mathcal{E}_x^2+\mathcal{E}_y^2-2\mathcal{E}_z^2\right)}{[4\pi\epsilon_0]m_1 m_2 \omega_1^2 \omega_2^2 R^3}
= \frac{\alpha_1\alpha_2~(\mathcal{E}_x^2+\mathcal{E}_y^2-2\mathcal{E}_z^2)}{[4\pi\epsilon_0]R^3}\ ,
\end{align}
where $\langle\bm{r}|n_x,n_y,n_z\rangle=\psi_{\{n_x,n_y,n_z\}}(\br)$ and $\langle n_x,n_y,n_z|\bm{r}\rangle$ is its complex conjugate. Equation~\eqref{ES-pert-mqm} is the same expression as the first term of Eq.~\eqref{nret-total-int-dissimilar}. 
To calculate the energy shift from the second-order perturbation,
\begin{align}
\label{2nd-pert-qm}
\Delta \mathscr{E}^{(2)}=\sum_{I\neq 0}\frac{\langle 0|V_{int}|I\rangle \langle I|V_{int}|0\rangle}{E_0-E_I}\ ,
\end{align}
we consider two cases regarding the states of unperturbed QDOs in the intermediate ket state $|I\rangle$\,: 
\begin{itemize}
\item \textbf{case $\bm{(i)}$}: one of the QDOs is in its excited state whereas the other one is in its ground state,
\item \textbf{case $\bm{(ii)}$}: both QDOs are in their excited states.
\end{itemize}
In the first case, at any instant of time, field-induced static dipole moment of just one of the atoms is involved in the interaction process between them, which yields the energy shift 
\begin{align}
\label{induction-mqm}
\!\!\Delta \mathscr{E}_{1}^{(2)}=&
-\!\sum_{\bm{n}\neq \bm{0}}
\frac{\Big[{}_2\langle \bm{0}|~{}_1\!\langle \bm{0}|~q_1q_2
	(\bm{r}_1\cdot\bm{r}_2 - 3z_1z_2) ~|\bm{n}\rangle_1~|\bm{0}\rangle_2\Big]^2}{[4\pi\epsilon_0]^2 R^6~\hbar\omega_1(n_x+n_y+n_z)}
\nonumber\\
&
\!\!-\!\sum_{\bm{m}\neq \bm{0}}
\frac{\Big[{}_2\langle \bm{0}|~{}_1\!\langle \bm{0}|~q_1q_2
	(\bm{r}_1\cdot\bm{r}_2 - 3z_1z_2) ~|\bm{0}\rangle_1~|\bm{m}\rangle_2\Big]^2}{[4\pi\epsilon_0]^2 R^6~\hbar\omega_2(m_x+m_y+m_z)}
\nonumber\\
=&-\frac{\alpha_1\alpha_2(\alpha_1+\alpha_2)~(\mathcal{E}_x^2+\mathcal{E}_y^2+4\mathcal{E}_z^2)}{2[4\pi\epsilon_0]^2R^6}
\ ,
\end{align} 
where $|\bm{n}\rangle=|n_x,n_y,n_z\rangle$ and $\bm{n}\neq \bm{0}$ means $\{n_x,n_y,n_z\}\neq\{0,0,0\}$. The interaction energy obtained in Eq.~\eqref{induction-mqm} is the same as the second term of Eq.~\eqref{nret-total-int-dissimilar}.
In the second case, where for each transition of the total system both QDOs are excited, the field-induced dipole moments do not contribute to the interaction. Therefore, Eq.~\eqref{2nd-pert-qm} yields the well-known dispersion energy
\begin{align}
\label{dispersion-mqm}
\Delta \mathscr{E}_{disp}^{(2)}\!\!&=\!
-\!\!\!\!\!\!\sum_{\bm{n},\bm{m}\neq \bm{0}}\!\!\!\!\!
\frac{\Big[{}_2\langle \bm{0}|~{}_1\!\langle \bm{0}|
	~q_1q_2(\bm{r}_1\cdot\bm{r}_2 - 3z_1z_2)~
	|\bm{n}\rangle_1~|\bm{m}\rangle_2\Big]^2}
	{[4\pi\epsilon_0]^2 R^6~(E_{\bm{n0}}+E_{\bm{m0}})}
\nonumber\\
&=\!
-\frac{3\hbar\omega_1\omega_2\alpha_1\alpha_2}{2[4\pi\epsilon_0]^2(\omega_1+\omega_2)R^6}\ ,
\end{align}
where we used the relations $E_{\bm{n0}}=\hbar\omega_1(n_x+n_y+n_z)$ and $E_{\bm{m0}}=\hbar\omega_2(m_x+m_y+m_z)$. The energy shift of Eq.~\eqref{dispersion-mqm} is the same as the third term of the nonretarded interaction energy given by Eq.~\eqref{nret-total-int-dissimilar}. 

Although the results obtained from the perturbation theory are approximate, they deliver all the leading contributions to the interaction energy. Moreover, this approach is more intuitive compared with the exact solution of the Schr\"odinger equation in terms of distinguishing the dipole moments involved in each contribution to the interaction energy between atoms/molecules. On the other hand, the diagonalization of the Hamiltonian in Eq.~\eqref{nret-Hamiltonian-x5} provides one with a more complete description of the effects of self-consistent electric fields.

In summary, the Rayleigh-Schr\"odinger perturbation theory considered in this section allowed us  to confirm the leading-order results obtained from exact diagonalization as coming from the first two orders of perturbation theory for two dipole-coupled QDOs. So far, all the derived energy terms correspond to the nonretarded regime of the interaction. However, for large interatomic separations in comparison to characteristic wavelengths of atomic transitions, the effect of retardation has to be taken into account. This implies that interactions are no longer instantaneous. This task can be accomplished by making use of a field-theoretical formalism, where the interaction between atoms occurs via exchanging photons. In the next section, we employ perturbation theory in the framework of microscopic QED, to investigate the effect of retardation on the interactions that were obtained so far in Sections III and VI.

\section{Perturbation theory in microscopic quantum electrodynamics}
Within the multipolar-coupling formalism of QED, interactions between atoms occur through their coupling to the fluctuating vacuum radiation field via their electric dipole/multipole moments, whereas any direct instantaneous coupling between atoms is eliminated. 
Therefore, for a system of two QDOs in presence of the vacuum radiation field as well as the external static electric field $\be$, the total Hamiltonian consists of the Hamiltonians of noninteracting QDOs and fields plus fields-QDOs coupling terms.
Similar to the derivation performed in the previous section, we consider the total unperturbed system as the system of two non-interacting QDOs that are already coupled to the external field via their electric dipole moments.
However, in contrast to the QM framework, here the perturbation occurs solely due to the coupling of the QDOs to the vacuum radiation field. Thus, in the  total Hamiltonian, $H=H^{(0)}+H_{\rm int}$, the Hamiltonian of the unperturbed system reads as 
\begin{align}
\label{nonperturbed-H}
H^{(0)}&=H_{\rm rad}+H_1^{(0)}+H_2^{(0)}\nonumber\\
&=H_{\rm rad}+\sum_{i=1,2}\left[\frac{\bm{p}_i^2}{2m}+\frac{1}{2}m\omega^2\br_i^2-\bm{\mu}_i\cdot\bm{\mathcal{E}}\right]
\ ,
\end{align}
where $\bm{\mu}_i=q_i\bm{r}_i$ is the electric dipole moment operator of the $i$th QDO, 
and $H_i^{(0)}$ is the Hamiltonian of an \emph{unperturbed QDO}, \textit{i.e.}~a QDO in the external field, with the eigenfunctions and energy eigenvalues given by Eqs.~\eqref{QDO-in-exF}.
The perturbation is given by
\begin{align}
\label{H_int_QED}
H_{\rm int}=&-\frac{1}{\epsilon_0}\bm{\mu}_1\cdot\bm{D}_\perp(\br_1)-\frac{1}{\epsilon_0}\bm{\mu}_2\cdot\bm{D}_\perp(\br_2)\ ,
\end{align}
where $\bm{D}_{\perp}$ is the transverse component of the vacuum displacement radiation field
\begin{align}
\bd_\perp(\br)=i\textstyle\sum\limits_{\bk,\lambda}\sqrt{\tfrac{\hbar c k \epsilon_0}{2 V}} \left(\eh_{\bk\lambda} a_{\bk\lambda}^{ }\ex^{i\bk\cdot\br}-\bar{\eh}_{\bk\lambda} a_{\bk\lambda}^\dagger\ex^{-i\bk\cdot\br}\right).
\end{align}
Here, $a_{\bk\lambda}$ and $a_{\bk\lambda}^\dagger$ are annihilation and creation operators of a vacuum-field mode with the wave vector $\bk$ and electric polarization vectors\ $\eh_{\bk\lambda}$ and $\bar{\eh}_{\bk\lambda}$, respectively. They obey the bosonic commutation relations
\begin{align}
[a_{\bk\lambda}\, , a_{\bk'\lambda'}^\dagger]=\delta_{\bk\bk'}\delta_{\lambda\lambda'}\, , \  
[a_{\bk\lambda}\, , a_{\bk'\lambda'}]=[a_{\bk\lambda}^\dagger\, , a_{\bk'\lambda'}^\dagger]=0\ .
\end{align}

The ground state ket vector of the total unperturbed system is given by the product state
\begin{align}
|0\rangle=|0,0,0\rangle_1~|0,0,0\rangle_2~|\{0\}\rangle\ ,
\end{align}
where $|\{0\}\rangle$ is the ground state of the vacuum radiation field and 
$\langle\bm{r}_i|0,0,0\rangle_i=\psi_{\{0,0,0\}}(\br_i)$ with $\psi(\br)$ given by Eq.~\eqref{QDO-in-exF1}.
The excited states of the total unperturbed system can be defined similarly. Then, making use of these states and the matrix elements of dipole moments, we perform QED perturbation-theory derivation, in order to obtain the interaction energy for the two QDOs, as consisting of contributions from different orders of corrections to the total energy of the unperturbed system.

The 1st- and the 3rd-orders of perturbation provide vanishing contributions because of the creation and annihilation operators of the radiation field sandwiched between two identical states of the vacuum, $|\{0\}\rangle$.

\begin{figure}[t!]
	\includegraphics[width=0.7\linewidth]{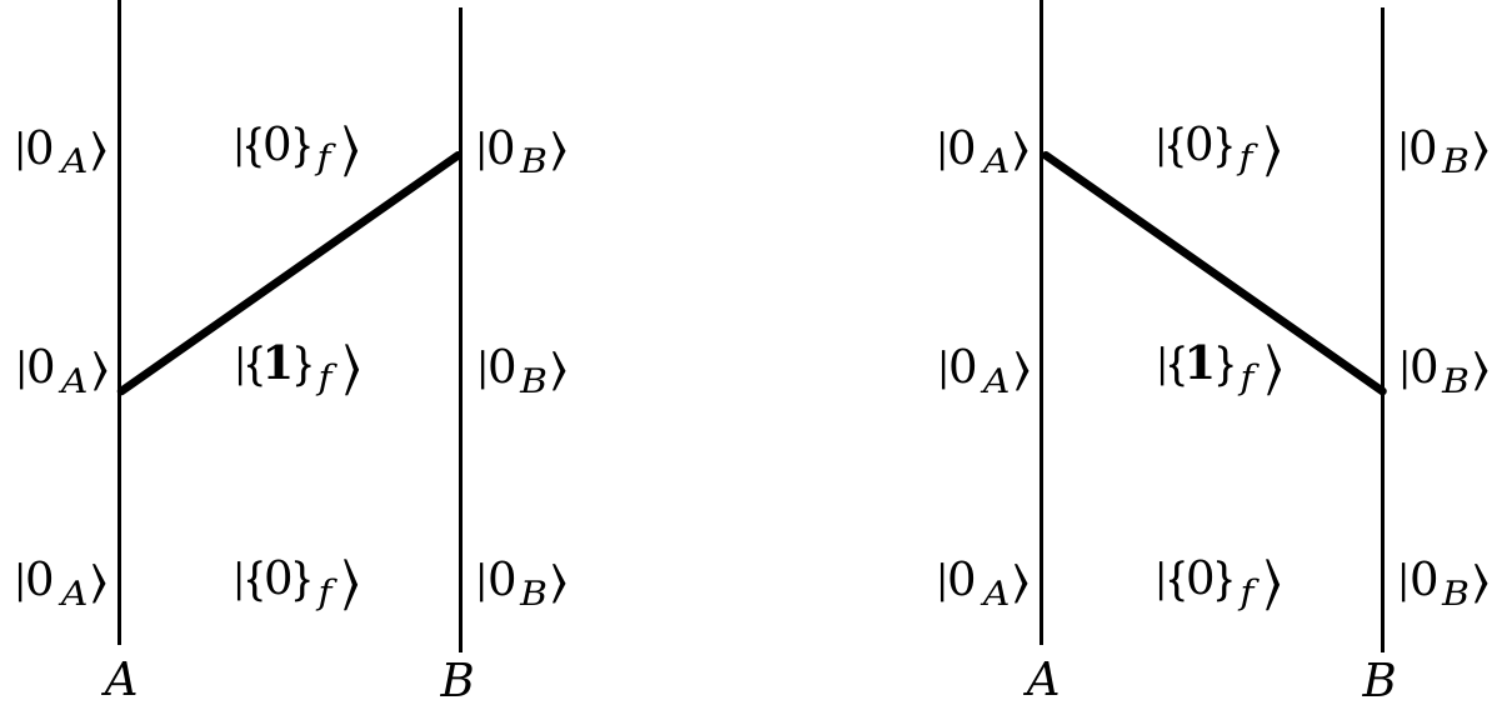}
	\caption{Two Feynman diagrams correspond to the coupling between static field-induced dipole moments of atoms and the vacuum field. The vertical solid lines are universal time lines.}
	\label{fig:diagramES}
\end{figure}

The non-vanishing terms from the second order, 
\begin{align}
\label{2ndpert}
\mathscr{E}^{(2)}=\sum_{I\neq 0}\frac{\langle 0|H_{\rm int}|I\rangle \langle I|H_{\rm int}|0\rangle}{E_0-E_I}\ ,
\end{align}
arise when the radiation field is excited with a single photon in the intermediate state $|I\rangle$. For the atomic part of $|I\rangle$, there are two possibilities that result in non-vanishing energy shifts which we consider separately: $(1)$ both unperturbed QDOs are in their ground states, $(2)$  one 
of them is excited, whereas the other one is in its ground state. 
In the former case, where the intermediate state $|I\rangle$ is defined as $|0,0,0\rangle_1|0,0,0\rangle_2|\bm{1}_{k\lambda}\rangle$,  the interaction between the atoms happens in two steps. 
First, one of the atoms interacts with the radiation field via its static field-induced dipole and emits a photon. 
Hence, the total system, which was initially in its ground state, is promoted to the excited state $|I\rangle$. 
At the second step, the other atom similarly interacts with the radiation field via its static field-induced dipole and absorbs the photon that was emitted at the first step. 
The second transition brings the total system back to the ground state. This procedure is equivalent to a sum over two distinct Feynman diagrams illustrated in Fig.~\ref{fig:diagramES}. 
They look similar to the diagrams corresponding to the interaction between molecules with permanent electric dipole moments~\cite{Craig1994}. 
The similarity suggests that the interaction energy stemming from the described mechanism corresponds to electrostatic interactions. 
This point as well as the origin of other contributions to the total interaction energy will be discussed in more detail within the next section, based on a transparent physical picture of the interactions provided by stochastic electrodynamics.

The sum in Eq.~\eqref{2ndpert} after removing self-energies, $\Delta \mathscr{E}^{(2)}=\mathscr{E}^{(2)}(R)-\mathscr{E}^{(2)}(\infty)$, reduces to
\begin{align}
\hspace{-0.2cm}
\Delta \mathscr{E}^{(2)}=-\sum_{\bm{k}\lambda}\frac{\mathcal{E}_i\mathcal{E}_j e_{\bk\lambda}^{(i)} e_{\bk\lambda}^{(j)}} {2V\epsilon_0}
\left[2\alpha_1\alpha_2\cos(k_zR)\right],
\hspace{-0.3cm}
\end{align}
where the repeated indices $i$ and $j$ imply summation over Cartesian components, $\{i,j\}=\{x,y,z\}$. Replacing the sum over $\bm{k}$ with a 3D integral, $\sum\rightarrow \frac{V}{8\pi^3}\int$, and summing over polarization of the radiation field yields the interaction energy 
\begin{align}
\label{es-qed1}
\Delta \mathscr{E}^{(2)}\!=\!-\frac{\alpha_1\alpha_2}{8\pi^3\epsilon_0} \mathcal{E}_i\mathcal{E}_j\!\!
\!\int\!\! d^3\bm{k}\,\cos(k_zR) (\delta_{ij}-\hat{k}_i\hat{k}_j)\ ,
\end{align}
with $\hat{k}_i={k_i}/{k}$.
Transforming this integral to the spherical coordinate system and performing the angular integration using the relation 
\begin{align}
\label{int_F(ij)}
\!\!\iint (\delta_{ij}-\hat{k}_i\hat{k}_j)\ex^{\pm i\bm{k}\cdot\bm{R}}\,\sin\theta d\theta\, d\varphi=4\pi {\rm Im}[F_{ij}(kR)]\, ,
\end{align}
with 
\begin{align}
\label{F(ij)}
F_{ij}(kR)=\Big[(\delta_{ij}-\hat{R}_i\hat{R}_j)\frac{1}{kR}&\\ \nonumber +(\delta_{ij}-3\hat{R}_i\hat{R}_j)&(\frac{i}{k^2R^2}-\frac{1}{k^3R^3})\Big]\ex^{ikR}\ ,
\end{align}
we arrive at
\begin{align}
\Delta \mathscr{E}^{(2)}\!=\!-\frac{\alpha_1\alpha_2}{8\pi^3\epsilon_0} \mathcal{E}_i\mathcal{E}_j\!\!
\!\int_0^\infty\!\!\!\! k^2
\Big[(\delta_{ij}-\hat{R}_i\hat{R}_j)\frac{\sin (kR)}{kR}\nonumber\\
+(\delta_{ij}-3\hat{R}_i\hat{R}_j)(\frac{\cos (kR)}{k^2R^2}-\frac{\sin(kR)}{k^3R^3})\Big]\,dk\ .
\end{align}
Carrying out the remaining integral, and keeping in mind that $\bm{R}=R\hat{\bm{z}}$, we obtain the interaction energy
\begin{align}
\label{ES-QED}
\Delta \mathscr{E}^{(2)}=\frac{\alpha_1\alpha_2(\mathcal{E}_x^2+\mathcal{E}_y^2-2 \mathcal{E}_z^2)}{4\pi\epsilon_0\, R^3}\ ,
\end{align}
as valid for any range of interatomic separation, $R$. The above expression reproduces the first term in Eq.~\eqref{nret-total-int-dissimilar} and hence is not affected by retardation. Since in this case both atoms are coupled to the vacuum field via their static field-induced dipoles, with the $R^{-3}$ distance dependence of the interaction energy given by Eq.~\eqref{ES-QED}, we call this contribution a field-induced electrostatic interaction.

For the second case, the intermediate state $|I\rangle$ corresponds to the situation when one of the QDOs is excited and the other one is in its ground state. 
For each transition of the total system to its excited state, one of the atoms emits a photon and then absorbs it by itself in the next downward transition, when the total system goes back to its ground state. 
Thus, for such series of transitions, there is no exchange of photons and hence no interaction between the atoms. 
Equation~\eqref{ES-QED} confirms our conclusion from the diagrams of Fig.~\ref{fig:diagramES}. 
Since, in the absence of the external field, there are no field-induced dipoles, the interaction energy, $\Delta \mathscr{E}^{(2)}$, vanishes similar to the case when the interaction occurs between molecules with no permanent electric dipoles. 
The power of the QDO model is that the effect of a static electric field clearly manifests as a shift in the center of oscillations of the Drude particle, which can be understood as a static polarization of the atom or molecule.

The third case, corresponding to the intermediate state $|I\rangle$ where both QDOs are excited, delivers only vanishing contributions due to the form of the interaction Hamiltonian given by Eq.~\eqref{H_int_QED}, which does not contain any direct coupling between the two oscillators. If both QDOs are simultaneously excited within the intermediate state $|I\rangle$, the resulting matrix element vanishes due to the orthogonality of the eigenstates of the oscillators,
\begin{align}
&\langle 0|H_{\rm int}|I\rangle=
- \epsilon_0^{-1} \times \\ \nonumber
&\langle\{0\}|{}_2\!\langle \bm{0}|{}_1\!\langle\bm{0}|
\big[\bm{\mu}_1\!\cdot\!\bm{D}_\perp(\br_1)\!+\!\bm{\mu}_2\!\cdot\!\bm{D}_\perp(\br_2)\big]
|\bm{n}\rangle_1|\bm{m}\rangle_2|\bm{1}_{\bm{k}\lambda}\rangle
\\ \nonumber
&= - \epsilon_0^{-1}
\Big[{}_1\!\langle \bm{0}|\bm{\mu}_1|\bm{n}\rangle_1
~{}_2\!\langle\bm{0}|\bm{m}\rangle_2\Big]
\cdot\langle\{0\}|\bm{D}_\perp(\br_1)|\bm{1}_{\bm{k}\lambda}\rangle
\\ \nonumber &
~~~\! - \epsilon_0^{-1}
\Big[{}_1\!\langle\bm{0}|\bm{n}\rangle_1~
{}_2\!\langle \bm{0}|\bm{\mu}_2|\bm{m}\rangle_2\Big]
\cdot\langle\{0\}|\bm{D}_\perp(\br_2)|\bm{1}_{\bm{k}\lambda}\rangle
=0\ \! ,
\end{align}
which gives no contributions to the interaction energy.

The next non-vanishing contribution to the interaction energy arises from the 4th-order perturbation theory~\rrr{\cite{Craig1994}}, 
\begin{align}
\label{4th-pert}
\hspace{-0.225cm} \mathscr{E}^{(4)}\!&=-\hspace{-0.4cm}\sum_{{I,I\!\!I,I\!\!I\!\!I}\neq 0}\hspace{-0.35cm}
\frac{
	\langle 0|H_{int}|I\!\!I\!\!I\rangle 
	\langle I\!\!I\!\!I|H_{int}|I\!\!I\rangle
	\langle I\!\!I|H_{int}|I\rangle 
	\langle I|H_{int}|0\rangle
}{(E_I-E_0)(E_{I\!\!I}-E_0)(E_{I\!\!I\!\!I}-E_0)}\nonumber\\
&\hspace{-0.55cm}+\hspace{-0.2cm}
\sum_{{I,I\!\!I}\neq 0}\hspace{-0.2cm}
\frac{
	\langle 0|H_{int}|I\!\!I\rangle 
	\langle I\!\!I|H_{int}|0\rangle
	\langle 0|H_{int}|I\rangle 
	\langle I|H_{int}|0\rangle
}{(E_I-E_0)^2(E_{I\!\!I}-E_0)}\ .
\end{align}
In the absence of the external field, similar to nonpolar molecules possessing no permanent dipole moments, the second term of Eq.~\eqref{4th-pert} does not contribute to the interaction energy. The first term contributes to $\Delta \mathscr{E}^{(4)}$ only from summing over
those combinations of intermediate states $|I\rangle,~|I\!\!I\rangle,$ 
and $|I\!\!I\!\!I\rangle$ that satisfy certain conditions, which are explained in the following. In the intermediate states 
$|I\rangle$ and $|I\!\!I\!\!I\rangle,$ the field must be in a single-photon excited state, while one of the atoms is excited and the other is in its ground state. 
Then for $|I\!\!I\rangle$ there are three possibilities that may result in finite contributions to the interaction energy: $(1$) the vacuum field is in a two-photon excitation state and both atoms are excited; $(2)$ the field is in a two-photon excitation state while both atoms are in their ground states; $(3)$ the field is in its ground state while both atoms are excited. Among all the possible combinations of such intermediate states, those that involve exchange of two virtual photons between the atoms lead to the dispersion interaction~\cite{Craig1994, Salam2009, Buhmann2013}. Within this picture, the interaction between the two atoms occurs through the coupling of their fluctuating electric dipole moments to the vacuum field.

In the presence of the external electric field, $\bm{\mathcal{E}}$, atoms become polarized possessing static field-induced dipole moments, $\bm{\mu}=\alpha\be$. Therefore, the coupling of atoms to the vacuum field may also happen through their static 
dipole moments, in addition to their fluctuating dipole moments. This additional possibility results in further contributions to the interaction energy. In what follows, we discuss such contributions by evaluating them from the first and the second terms of Eq.~\eqref{4th-pert}. This task is performed for three separate cases, listed in Table~\ref{tab:4thcases}, depending on the atomic behavior in the intermediate states $|I\rangle$, $|I\!\!I\rangle$, and $|I\!\!I\!\!I\rangle$.
\begin{table}[h!]
 \caption{Atomic behavior in the intermediate states, which appear within the 4th-order perturbation theory, Eq.~\eqref{4th-pert}.}
 \label{tab:4thcases}
 \begin{ruledtabular}
 \begin{tabular}{cc}
 ~\textbf{Case} & Atomic transitions in virtual states \\
 \hline
~$\bm{(c1)}$  & Both atoms do transitions \\
 & \\
~$\bm{(c2)}$ & One of the atoms remains in the ground \\
 & state, the other atom does transitions \\
 & \\
~$\bm{(c3)}$ & None of the two atoms do transitions \\
 \end{tabular}
 \end{ruledtabular}
 \end{table}

\begin{figure}[h!]
	\includegraphics[width=0.65\linewidth]{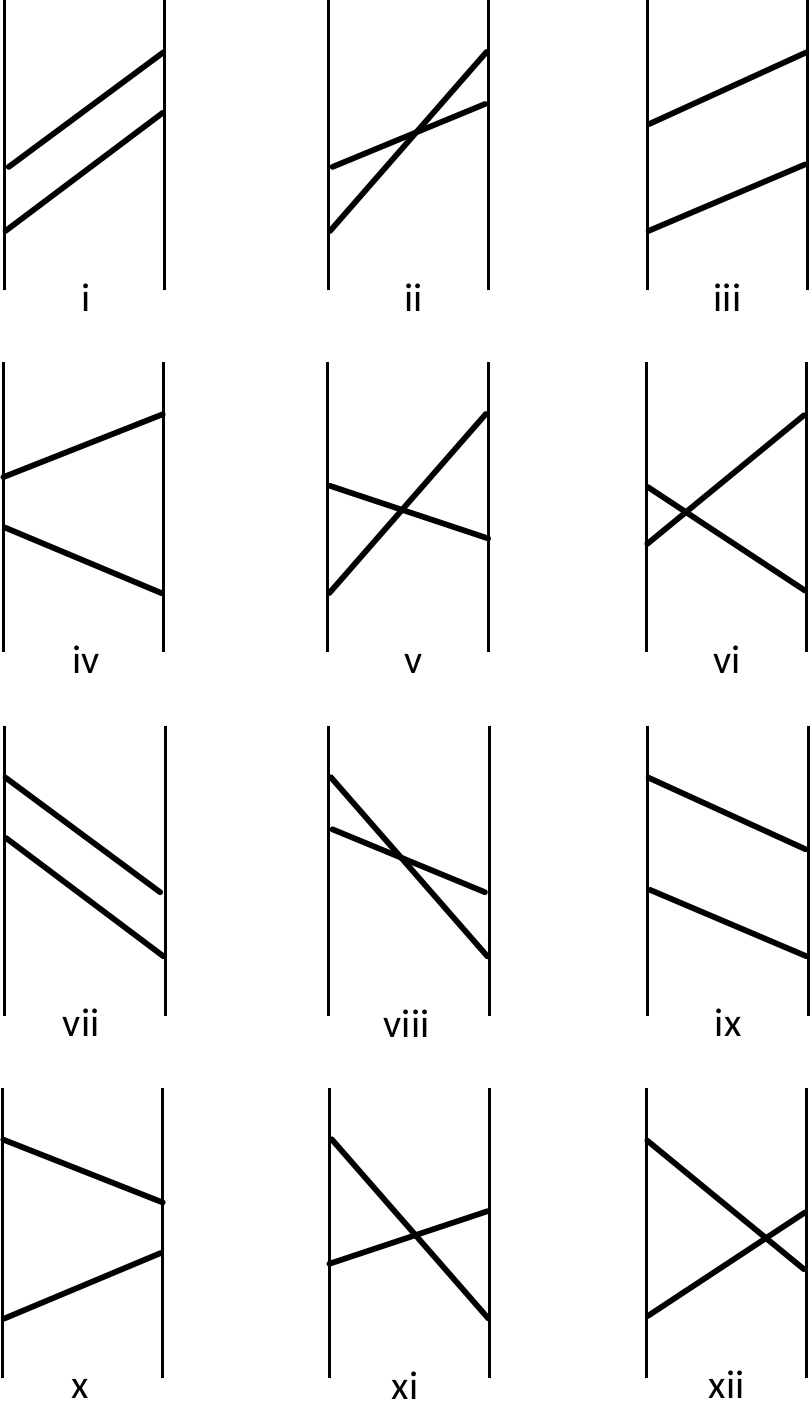}
	\caption{The twelve diagrams contributing to the first term of 
	the 4th-order energy correction given by Eq.~\eqref{4th-pert}. This figure follows Fig.~7.5 of Ref.~\cite{Craig1994}.}
	\label{fig:diagrams}
\end{figure}

\textbf{Case $\bm{(c1)}$} is similar to the situation of nonpolar species, as was already discussed above. Therefore, the resulting interaction energy 
in this case should be the same as the dispersion energy of two coupled nonpolar atoms in absence of any external field. The calculation of this energy shift follows the standard procedure presented in Refs.~\cite{Craig1994, Salam2009}. As explained there, the dispersion interaction between two atoms arises due to the exchange of a pair of virtual photons. For instance, such an exchange may happen through the following steps:
\begin{figure}[h!]
	\includegraphics[width=0.35\linewidth]{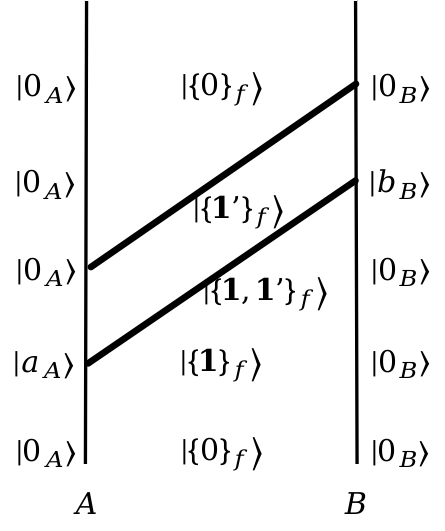}
	\caption{An example of exchange of two virtual photons between two atoms $A$ and $B$.}
	\label{fig:diagram1}
\end{figure} 
\begin{enumerate}
	\item Atom $A$ goes to an excited state $|a\rangle$ and emits a virtual photon $\bm{1}_{\bm{k}\lambda}$\,, while atom $B$ remains in its ground state: $|0_A,0_B,\{0\}_f\rangle \longrightarrow |a_A,0_B,\{\bm{1}_{\bm{k}\lambda}\}_f\rangle$\,;
	\item Atom $A$ gets de-excited and emits another photon, $\bm{1}_{\bm{k}'\lambda'}$\,, but atom $B$ is still 
	in its ground state: \\
	$|a_A,0_B,\{\bm{1}_{\bm{k}\lambda}\}_f\rangle \longrightarrow |0_A,0_B,\{\bm{1}_{\bm{k}\lambda},\bm{1}_{\bm{k}'\lambda'}\}_f\rangle$\,;
	\item Atom $B$ absorbs one of the photons and transits to an excited state $|b\rangle$, while $A$ remains unchanged: 
	$|0_A,0_B,\{\bm{1}_{\bm{k}\lambda},\bm{1}_{\bm{k}'\lambda'}\}_f\rangle \longrightarrow |0_A,b_B,\{\bm{1}_{\bm{k}'\lambda'}\}_f\rangle$\,;
	\item Atom $B$ absorbs the other photon and goes back to its ground state, while $A$ remains unchanged:\\ 
	$|0_A,b_B,\{\bm{1}_{\bm{k}'\lambda'}\}_f\rangle \longrightarrow |0_A,0_B,\{0\}_f\rangle$\,.
\end{enumerate}
This four-step procedure is illustrated in Fig.~\ref{fig:diagram1}.
In total, there are twelve distinct diagrams representing all possible combinations of atomic and field states. They correspond to the exchange of a pair of virtual photons between the two atoms, as shown in Fig.~\ref{fig:diagrams}.
All the twelve diagrams give the same numerators in the first term of Eq.~\eqref{4th-pert} 
but different denominators (see Table~\ref{tab:denominators}).
\begin{table}[h!]
   \caption{Denominators of the first term in Eq.~\eqref{4th-pert} related to diagrams of Fig.~\ref{fig:diagrams}. Our table follows Table~7.1 of Ref.~\cite{Craig1994}. $E_{a0}=E_{a}-E_{0}$ and $E_{b0}=E_{b}-E_{0}$ denote excitation energies of atoms $A$ and $B$ to their excited states $|a\rangle$ and $|b\rangle$, respectively.}
   \label{tab:denominators}
 \begin{ruledtabular}
 \begin{tabular}{cc}
 ~Diagram & Denominator \\
 \hline
 ~(i)  & $(E_{b0}+\hbar c k)(\hbar c k + \hbar c k')(E_{a0}+\hbar c k')$~ \\
 ~(ii)  & $(E_{b0}+\hbar c k')(\hbar c k + \hbar c k')(E_{a0}+\hbar c k')$~ \\
 ~(iii)  & $(E_{b0}+\hbar c k)(E_{a0} + E_{b0})(E_{a0} + \hbar c k')$~ \\
 ~(iv)  & $(E_{b0}+\hbar c k)(E_{a0} + E_{b0})(E_{b0} + \hbar c k')$~ \\
 ~(v)  & $(E_{b0}+\hbar c k')(E_{a0}+E_{b0}+\hbar c k + \hbar c k')(E_{a0}+\hbar c k')$~ \\
 ~(vi)  & $(E_{b0}+\hbar c k')(E_{a0}+E_{b0}+\hbar c k + \hbar c k')(E_{b0}+\hbar c k)$~ \\
 ~(vii)  & $(E_{a0}+\hbar c k)(\hbar c k + \hbar c k')(E_{b0}+\hbar c k')$~ \\
 ~(viii)  & $(E_{a0}+\hbar c k)(\hbar c k + \hbar c k')(E_{b0}+\hbar c k)$~ \\
 ~(ix)  & $(E_{a0}+\hbar c k)(E_{a0} + E_{b0})
 (E_{b0}+\hbar c k')$~ \\
 ~(x)  & $(E_{a0}+\hbar c k)(E_{a0} + E_{b0})(E_{a0}+\hbar c k')$~ \\
 ~(xi)  & $(E_{a0}+\hbar c k)(E_{a0} + E_{b0} + \hbar c k + \hbar c k')(E_{b0}+\hbar c k)$~ \\
 ~(xii)  & $(E_{a0}+\hbar c k)(E_{a0} + E_{b0} + \hbar c k + \hbar c k')(E_{a0}+\hbar c k')$~ \\
 \end{tabular}
 \end{ruledtabular}
 \end{table}

Summing over all these terms, as well as performing summations and integrals over all atomic states and modes of the vacuum field, yields the well-known London and Casimir-Polder dispersion energy~\cite{Craig1994,Salam2009},
\begin{subequations}
\label{dispersion-qed}
\begin{align}
&\Delta \mathscr{E}^{(4)}_1=\Delta \mathscr{E}_{_{\rm L}}=-\frac{3\alpha_1 \alpha_2 ~ \omega_1 \omega_2 \hbar }{2 [4\pi\epsilon_0]^2 (\omega_1+\omega_2) R^6}\ ,\\
&\Delta \mathscr{E}^{(4)}_1=\Delta \mathscr{E}_{_{\rm CP}}=-\frac{23 \hbar c ~\alpha_1\alpha_2}{[4\pi\epsilon_0]^2 4\pi R^7}\ ,
\end{align}
\end{subequations}
for nonretarded and retarded regimes, respectively.
As mentioned above, the second term in Eq.~\eqref{4th-pert} does not contribute to the interaction energy in case $(c1)$.

In \textbf{case~$\bm{(c2)}$} of Table~\ref{tab:4thcases}, one of the atoms is coupled to the vacuum field via its static field-induced dipole, while the other atom may couple to the vacuum field via its fluctuating dipole moment. 
However, for the interaction between two atoms, they should exchange a pair of virtual photons. 
Hence, the interactions should be described by the same diagrams as shown in Fig.~\ref{fig:diagrams}. 
The denominators of the first term in Eq.~\eqref{4th-pert} corresponding to the twelve diagrams are almost the same as those given in Table~\ref{tab:denominators}. 
The only difference is that the atomic transition energy must be replaced with zero for the atom that is coupled to the vacuum field via its field-induced dipole moment and does not undergo any transition ($\hbar ck_{a0}=0$ or $\hbar ck_{b0}=0$). 
Therefore, in case $(c2)$ the 4th-order interaction energy is given by the sum of two energy shifts $\Delta \mathscr{E}^{(4)}_{2}=\sum_{n=1}^2[\Delta \mathscr{E}_{2}^{(4)}]_n$ where $n$ denotes the atom that remains in its ground state during the interaction.
Evaluating the numerator and denominator of the first term in Eq.~\eqref{4th-pert} for each diagram and summing over all the
twelve diagrams yields 
\begin{align}
\label{DeltaE4-ret-2}
[\Delta {\mathscr{E}^{(4)}_{2}}]_1\!\!=\!&-\!
\frac{\alpha_1^2\mathcal{E}_i\mathcal{E}_j}{V^2\epsilon_0^2 ~\hbar c}
\sum_{\bk,\bk'} \sum_{\lambda,\lambda'} \sum_b
\hat{e}_{\bk\lambda}^{(i)}\,
\bar{\hat{e}}_{\bk\lambda}^{(i')}\,
\hat{e}_{\bk' \lambda'}^{(j)}\,
\bar{\hat{e}}_{\bk' \lambda'}^{(j')} \nonumber\\
&~\quad\times\frac{\mu_{i'}^{0b}\mu_{j'}^{b0}}{k_{b0}}
\left(\frac{1}{k+k'}-\frac{1}{k-k'}\right)
k'\ex^{i(\bm{k}+\bm{k}')\cdot\bm{R}}
\nonumber\\
&\hspace{-1.5cm}=
-\frac{\alpha_1^2\mathcal{E}_i\mathcal{E}_j}{64\pi^6 \epsilon_0^2 ~\hbar c}
\!\sum_{b}
\frac{\mu_{i'}^{0b}\mu_{j'}^{b0}}{k_{b0}}
\!\!\iint\!\! k'(\delta_{ii'}-k_ik_{i'})(\delta_{jj'}-k'_jk'_{j'})\nonumber\\
&\hspace{-0.25cm}\times\ex^{i(\bm{k}+\bm{k}')\cdot\bm{R}}
\left(\frac{1}{k+k'}-\frac{1}{k-k'}\right)d^3k\, d^3k'\ ,
\end{align}
where $\{i,j,i',j'\}=\{x,y,z\}$ and sums over repeated indices are implied. The sum $\sum_b$ runs over all atomic states of the second atom with 
$\hbar ck_{b0}=E_{b}-E_0$. Here, to move from the r.h.s.~of the first equality to the r.h.s.~of the second equality, we performed summations over the vacuum-field polarization and replaced sums over $\bm{k}$ and $\bm{k}'$ with the related integrals. By transforming the latter to spherical coordinates, performing angular integration, and evaluating the integral over $k'$, 
like in Refs.~\cite{Craig1994, Salam2009}, the energy shift $[\Delta \mathscr{E}^{(4)}_2]_1$ becomes
\begin{align}
\label{DeltaE4-ret-3}
[\Delta \mathscr{E}^{(4)}_2]_1=&-\frac{1}{4\pi^3\epsilon_0^2}
\alpha_1^2\mathcal{E}_i\mathcal{E}_j
\left(\!\sum_s\frac{\mu_{i'}^{0b}\mu_{j'}^{b0}}{\hbar c k_{b0}}\!\right)\nonumber\\
&\times\int_0^\infty\!\! k^5 \mathrm{Re}[F_{jj'}(kR)]\mathrm{Im}[F_{ii'}(kR)]\, dk\ .
\end{align}

Considering Eq.~\eqref{xmatel}, one can show that only the first three excited states ($|100\rangle, ~|010\rangle$, and $|001\rangle$) of the second QDO contribute to the sum over atomic states. Then, using the result of Eq.~\eqref{xmatel}, we have
\begin{widetext}
\begin{align}
\label{DeltaE4-ret-4}
[\Delta \mathscr{E}^{(4)}_2]_1&\!=\!\frac{-1}{4\pi^3\epsilon_0^2}\left\{
\frac{\alpha_1^2(\mathcal{E}_x^2+\mathcal{E}_y^2)}{\hbar\omega_2}\left(q_2\sqrt{\tfrac{\hbar}{2m_2\omega_2}}\right)^2\right.\!\!\!\!\!
\int_0^\infty\!\!\!\! k^5\!
\left[\frac{\sin (2 k R)}{2 k^2 R^2}\!
+\!\frac{\cos (2 k R)}{k^3 R^3}\!
-\!\frac{3 \sin (2 k R)}{2 k^4 R^4}\!
-\!\frac{\cos (2 k R)}{k^5 R^5}\!
+\!\frac{\sin (2 k R)}{2 k^6 R^6}\right]\!dk
\nonumber\\
&\left.+\frac{\alpha_1^2\mathcal{E}_z^2}{\hbar\omega_2}\!\left(q_2\sqrt{\tfrac{\hbar}{2m_2\omega_2}}\right)^2\!\!\!\!
\int_0^\infty\!\!\!\! k^5\!
\left[-\frac{2 \sin (2 k R)}{k^4 R^4}\!
-\!\frac{4 \cos (2 k R)}{k^5 R^5}\!
+\!\frac{2 \sin (2 k R)}{k^6 R^6}\right]\!dk\!
\right\}\! = -\frac{\alpha_1^2 \alpha_2~(\mathcal{E}_x^2+\mathcal{E}_y^2+4\mathcal{E}_z^2)}{2 [4\pi\epsilon_0]^2 R^6}\ .
\end{align}
\end{widetext}
Here, the QDOs are assumed to be isotropic and the integrals are taken using standard integration techniques without any specific assumption about $R$, which makes Eq.~\eqref{DeltaE4-ret-4} valid for any range of interatomic separation. The term $[\Delta \mathscr{E}^{(4)}_2]_2$ can be similarly obtained. 
In case~$(c2)$, the second term of Eq.~\eqref{4th-pert} does not contribute to the interaction energy between the QDOs since the resulting energy shift has no distance-dependent part. 
Therefore, the total energy shift from case~$(c2)$ is given by
\begin{equation}
\label{induction-qed}
\Delta \mathscr{E}_{2}^{(4)}=
-\frac{\alpha_1 \alpha_2(\alpha_1+\alpha_2)~(\mathcal{E}_x^2+\mathcal{E}_y^2+4\mathcal{E}_z^2)}{2 [4\pi\epsilon_0]^2 R^6}\ ,
\end{equation}
which is the same as the second term of Eq.~\eqref{nret-total-int-dissimilar}. 
Since this interaction results from the coupling of the two atoms to the vacuum field, one by its static field-induced dipole moment and the other one by its fluctuating dipole moment, it is apparent that this term corresponds to polarization (induction) interactions. However, here the static dipoles are initially induced by the applied static electric field. Hence, we relate the energy given by Eq.~\eqref{induction-qed} to a field-induced polarization interaction.

In \textbf{Case}~$\bm{(c3)}$ of Table~\ref{tab:4thcases}, each of the two terms of Eq.~\eqref{4th-pert} 
provides contributions to the interaction energy.
Among the twelve diagrams of Fig.~\ref{fig:diagrams}, four of them (iii, iv, ix, and x) contribute to the second term. The other eight diagrams contribute to the first term of Eq.~\eqref{4th-pert}. They correspond to expressions with similar numerators but different denominators. Denoting the latter by $D_n$, there are three distinct cases for the eight diagrams:
\begin{align}
& D_{\rm i}=D_{\rm vi}=D_{\rm vii}=D_{\rm xii}=\hbar^3 c^3 k k'(k+k')\ ,\\
\nonumber & D_{\rm ii}=D_{\rm v}=\hbar^3 c^3 {k'}^2(k+k') , \, 
D_{\rm viii}=D_{\rm xi}=\hbar^3 c^3 {k}^2(k+k')\, .
\end{align}
Summing over inverse of these denominators yields
\begin{equation}
\sum_n \frac{1}{D_n}=\frac{2(k+k')}{\hbar^3 c^3 k^2 {k'}^2}\ .
\end{equation}
Therefore, for the first term of the 4th-order energy correction, after carrying out the sum over the radiation field polarization and replacing the sums over the wave vectors with integrals, we have
\begin{align}
[\Delta \mathscr{E}^{(4)}_{3}]_1=-\frac{\alpha_1^2\alpha_2^2\,\mathcal{E}_i\mathcal{E}_j\mathcal{E}_{i'}\mathcal{E}_{j'}}{(2\pi)^6\, 2\epsilon_0^2\,\hbar c}
\iint d^3\bm{k}~d^3\bm{k}'\bigg[\frac{k+k'}{k\,k'}\nonumber\\
(\delta_{ii'}-\hat{k}_i\hat{k}_{i'})\ex^{i\bm{k}\cdot\bm{R}}\,
(\delta_{jj'}-\hat{k}\rrr{'}_{\!\!j}\hat{k}\rrr{'}_{\!\!j'})\ex^{i\bm{k}\rrr{'}\cdot\bm{R}}\bigg]\ .
\end{align}
Similar to the previous cases, transforming the
integrals to the spherical coordinate system and taking the angular integrals using Eqs.~\eqref{int_F(ij)} and \eqref{F(ij)}, gives us
\begin{align}
&[\Delta \mathscr{E}^{(4)}_{3}]_1=-\frac{\alpha_1^2\alpha_2^2\,\mathcal{E}_i\mathcal{E}_j\mathcal{E}_{i'}\mathcal{E}_{j'}}{8\pi^4\, \epsilon_0^2\,\hbar c} \\
\nonumber &\times\!\!\!\iint\! kk'(k+k'){\rm Im}[F_{ii'}(kR)] {\rm Im}[F_{jj'}(k'R)]\,dk dk'\ \!.
\!\!\!\!
\end{align}
Then, the integration over $k'$ results in
\begin{align}
[\Delta& \mathscr{E}^{(4)}_{3}]_1=-\frac{\alpha_1^2\alpha_2^2\,\mathcal{E}_i\mathcal{E}_j\mathcal{E}_{i'}\mathcal{E}_{j'}}{8\pi^4\, \epsilon_0^2\,\hbar c}
\int_0^\infty\! \bigg\{k\,{\rm Im}[F_{ii'}(kR)]\nonumber\\
&\times\left(2\hat{R}_j\hat{R}_{j'}\frac{k}{R^2}-(\delta_{jj'}-3\hat{R}_j\hat{R}_{j'})\frac{\pi}{2R^3}\right)\!\bigg\} dk\ .
\end{align}
In addition, the $k$--integral can be taken on making use of elementary integration techniques with no need for any specific assumption about the range of $R$. When replacing $\bm{R}$ by $R_z\hat{\bm{z}}$, we obtain
\begin{align}
\label{R5-qed}
[\Delta \mathscr{E}^{(4)}_{3}]_1=+\frac{4\alpha_1^2\alpha_2^2~\mathcal{E}_z^2(\mathcal{E}_x^2+\mathcal{E}_y^2-2\mathcal{E}_{z}^2)}{[4\pi \epsilon_0]^2\,(\hbar c\pi)\,R^5}\ .
\end{align}

The remaining four diagrams (iii, iv, ix, and x) equally contribute to the energy shift $[\Delta \mathscr{E}^{(4)}_{3}]_2$ resulting from the second term of Eq.~\eqref{4th-pert}. 
Performing steps similar to our derivation of
$[\Delta \mathscr{E}^{(4)}_{3}]_1$, one can obtain
$[\Delta \mathscr{E}^{(4)}_{3}]_2$. 
The two terms turn out to be $[\Delta \mathscr{E}^{(4)}_{3}]_2=-[\Delta \mathscr{E}^{(4)}_{3}]_1$, which consequently give no net contribution to the total interaction energy between the atoms when both are coupled to the radiation field via their static field-induced dipole moments. 
Thus, the interaction energies given by Eqs.~\eqref{dispersion-qed} and \eqref{induction-qed} are  the leading terms from the 4th-order perturbation theory.

The fact that the interaction energies of Eqs.~\eqref{ES-QED} and \eqref{induction-qed} are not affected by the retardation might seem to be surprising at the first glance. However, once the origin of these interactions is identified, their static behavior becomes understandable. Within the next section, we perform a derivation in the framework of stochastic electrodynamics and identify the origin of each contribution to the interaction energy obtained thus far.

\section{Stochastic electrodynamics}
Finally, we employ a semiclassical approach, mainly
developed by Boyer \cite{Boyer1969, Boyer1971, Boyer1972, Boyer1973}, to derive the interaction energy from classical electrodynamics with a classical random electromagnetic zero-point radiation field. Within this approach, the random radiation field, which is a classical equivalent of the vacuum fluctuating radiation field in QED, polarizes atoms. Then the induced random polarizations of nearby atoms interact through their electromagnetic fields obeying principles of classical electrodynamics. Here, we restrict our consideration to the retarded regime, where for large interatomic distances only low frequencies (large wavelengths) significantly contribute to the retarded interactions. The nonretarded case can be similarly considered following Ref.~\cite{Boyer1972}. The stochastic electrodynamics approach permits a straightforward identification of the different interaction terms with the electric fields that cause them, providing a minimal model to understand the origin of molecular interactions.

Let us consider a classical dipole oscillator with charge $q$, mass $m$, and characteristic frequency $\omega$. These parameters are again to be determined by the conditions of Eq.~\eqref{QDO-parameter}. In an electric field $\bm{E}(\bm{r},t)$, the equation of motion of such a classical counterpart of the QDO is given by~\cite{Boyer1971}
\begin{align}
\label{Eq_motion_SED1}
m\frac{d^2 \bm{r}}{dt^2}=-m\omega^2\bm{r}+q\bm{E}(\bm{r},t)+\tau\frac{d^3 \bm{r}}{dt^3}\ ,
\end{align}
where the last term corresponds to the radiation reaction. For each mode of the electric field $\bm{E}$ with frequency $\Omega$, the above equation reduces to
\begin{align}
\label{eq-motion-w}
-m\Omega^2\bm{r} = -m\omega^2 \bm{r}^2 + q \bm{E}_{\Omega}(\bm{r},t) + i\tau \Omega^3 \bm{r}\ .
\end{align}
Here, being interested in the retarded regime (large separation distances), we can assume that only modes with low frequencies contribute to the coupling between the two species. 
As discussed in Refs.~\cite{Boyer1971,Boyer1973}, this assumption is valid since electromagnetic waves with large $\Omega$ (short wavelengths) have destructive interference with the waves of adjacent frequencies due to slight phase shifts acquired at large distances. This effect of mutual cancellations for high-frequency modes leads to a situation when only waves with large wavelengths (compared to the separation distance) contribute to the interaction between the two species. 
Therefore, one can assume that the terms 
$\propto$ $\omega^2$ and $\omega^3$ are much smaller
than the two other terms in Eq.~\eqref{eq-motion-w}.
By neglecting such small terms for all modes of the field $\bm{E}(\bm{r},t)$, Eq.~\eqref{Eq_motion_SED1} reduces to
\begin{align}
m \omega^2 \bm{r} = q \bm{E}(\bm{r},t)\ .
\end{align}
Now replacing $q^2/m \omega^2$  with the static polarizability, 
$\alpha$, we obtain the oscillator dipole as
$\bm{\mu} \equiv q\, \bm{r} = \alpha \bm{E}(\bm{r},t)$, 
where $\bm{E}(\bm{r},t)$ is the total electric field at its position.

The energy of an electric dipole moment induced by an electric field in the same field is known from classical electrodynamics as given by $\mathscr{E}=-\frac{1}{2}\alpha\langle\bm{E}^2\rangle$, where the bracket indicates time-averaging. 
Here, we apply a static uniform electric field on top of the random zero-point radiation field. Consequently, the induced polarization of an oscillator has two parts each corresponding to one of the fields. We assume that the first oscillator is located at the origin,
$\bm{r}_1=(0,0,0)$, and we bring the second oscillator 
to the point $\bm{r}_2=(0,0,R)$ on the $z$
axis from its initial position $(0,0,+\infty)$.
The energy difference of the total system in these two configurations, 
$\Delta\mathscr{E}(R)=\mathscr{E}(R)-\mathscr{E}(\infty)$, is the interaction energy that we are looking for. The total electric field at the position of the second oscillator
is a vector sum of the four fields
\begin{equation}
\label{total-E-r2}
\bm{E}(\bm{r}_2,t)=\bm{E}_0(\bm{r}_2,t)+\bm{E}_{\mu_1}(\bm{r}_2,t)+\bm{\mathcal{E}}_{\mu_1}(\bm{r}_2)+\bm{\mathcal{E}}\ .
\end{equation}
Here, the letters $\bm{\mathcal{E}}$ and $\bm{E}$ denote the electrostatic and radiation fields, respectively, where
$\bm{E}_0(\bm{r}_2,t)$ is the random 
zero-point radiation field defined by~\cite{Boyer1975}
\begin{equation}
\label{zero-point-field}
\bm{E}_0(\bm{r},t)=\mathrm{Re}\!\sum_{\lambda=1}^{2}\!\int d^3k \tfrac{\bm{\epsilon}(\bm{k},\lambda)\mathfrak{h}(\bm{k},\lambda)}{\sqrt{4\pi\epsilon_0}} \ex^{i[\bm{k}\cdot\bm{r}-\ddd{\Omega} t+\theta(\bm{k},\lambda)]}\, .\!\!
\end{equation}
For each mode of the field, $\mathfrak{h}^2$ is the energy associated to that mode 
$\left(\mathfrak{h}^2(\bm{k},\lambda) = \hbar\, \Omega/2\pi^2\right)$, $\theta(\bm{k},\lambda)$
is a random phase ranging from $0$ 
to $2\pi$, $\bm{\epsilon}(\bm{k},\lambda)$ are orthogonal polarization unit vectors with 
$\bm{\epsilon}(\bm{k},\lambda)\cdot\bm{\epsilon}(\bm{k}',\lambda')=\delta_{\lambda\lambda'}$, and the sum runs over two possible polarizations. Then, $\bm{E}_{\mu_1}(\bm{r}_2,t)$ is a time-dependent field radiated from the oscillating dipole of the first oscillator 
induced by the zero-point radiation field. By analogy, $\bm{\mathcal{E}}_{\mu_1}(\bm{r}_2)$ is the electric field of 
the static dipole of the first oscillator
induced by the uniform electric field $\bm{\mathcal{E}}$. 
The electric fields of static and oscillating dipole moments are given by~\cite{Jackson1998}
\begin{equation}
\label{E-p-static-field}
\bm{\mathcal{E}}_{\mu}(\bm{r})=\frac{1}{4\pi\epsilon_0}\frac{3\bm{n}(\bm{\mu}\cdot\bm{n})-\bm{\mu}}{r^3}\ ,
\end{equation}
and
\begin{equation}
\label{E-p-rad-field}
\hspace{-0.15cm}
\bm{E}_{\mu}(\bm{r},t)\!=\!\mathrm{Re}\!\left[\!\left(\tfrac{k^2(\bm{n}\times\bm{\mu})\times\bm{n}}{4\pi\epsilon_0\ r}
+\bm{\mathcal{E}}_{\mu}(\bm{r})\left(1-ikr\right)\right)\ex^{ikr}\!\right] ,
\end{equation}
respectively.
Thus, the electromagnetic energy of the second oscillator, located
at $\bm{r}_2=(0,0,R)$ and possessing a static polarizability $\alpha_2$, in the presence of the total electric field given by Eq.~\eqref{total-E-r2} 
can be obtained in the lowest order of coupling as
\begin{align}
\label{Energy-in-field}
\mathscr{E}_{2}(R)\!&=\!-\frac{1}{2}\alpha_2\left\langle\bm{E}^2(\bm{R},t)\right\rangle \\ \nonumber
&=\!-\frac{\alpha_2}{2}\left\langle\!\left[\bm{E}_0(\bm{R},t)\!+\!\bm{E}_{\mu_1}\!(\bm{R},t)\!+\!\bm{\mathcal{E}}_{\mu_1}\!(\bm{R})\!+\!\bm{\mathcal{E}}\right]^2\!\right\rangle .
\end{align}
After subtracting the oscillator self-energy at $R \to +\infty$ from 
Eq.~\eqref{Energy-in-field}, we arrive at
\begin{align}
\label{ret-deltaE-t0}
\Delta\mathscr{E}_2=-\alpha_2\bigg[
&\Big\langle\bm{\mathcal{E}}\cdot\bm{\mathcal{E}}_{\mu_1}(\bm{R})\Big\rangle
+\frac{1}{2}\Big\langle\bm{\mathcal{E}}_{\mu_1}(\bm{R})\cdot\bm{\mathcal{E}}_{\mu_1}(\bm{R})\Big\rangle
\nonumber\\
+&\Big\langle\bm{E}_0(\bm{R},t)\cdot\bm{E}_{\mu_1}(\bm{R},t)\Big\rangle
+\Big\langle\bm{\mathcal{E}}\cdot\bm{E}_{\mu_1}(\bm{R},t)\Big\rangle
\nonumber\\
+&\Big\langle\bm{\mathcal{E}}\cdot\bm{E}_0(\bm{R},t)\Big\rangle
+\Big\langle\bm{\mathcal{E}}_{\mu_1}(\bm{R})\cdot\bm{E}_{\mu_1}(\bm{R},t)\Big\rangle
\nonumber\\
+&\Big\langle\bm{\mathcal{E}}_{\mu_1}(\bm{R})\cdot\bm{E}_0(\bm{R},t)\Big\rangle
\bigg]\ .
\end{align}

First, we perform averaging over time and random phase by making use of the following relations
\begin{align}
\label{averaging-1}
&\left\langle \cos[-\Omega\, t+\theta_{\bm{k}\lambda}] \right\rangle =
\left\langle \sin[-\Omega\, t+\theta_{\bm{k}\lambda}] \right\rangle = 0\ ,
\\ \nonumber
&\left\langle \sin[-\Omega\, t+\theta_{\bm{k}\lambda}] \cos[-\Omega' t+\theta_{\bm{k}'\lambda'}]\right\rangle  = 0\ ,
\\ \nonumber
&\left\langle \cos[-\Omega\, t+\theta_{\bm{k}\lambda}] \cos[-\Omega' t+\theta_{\bm{k}'\lambda'}] \right\rangle =
\tfrac{1}{2}\, \delta_{\lambda\lambda'}\, \delta_{\bm{k}\bm{k}'}\ ,
\hspace{-0.2cm}
\\ \nonumber
&\left\langle \sin[-\Omega\, t+\theta_{\bm{k}\lambda}] \sin[-\Omega' t+\theta_{\bm{k}'\lambda'}] \right\rangle =
\tfrac{1}{2}\, \delta_{\lambda\lambda'}\, \delta_{\bm{k}\bm{k}'}\ ,
\hspace{-0.2cm}
\end{align}
where $\theta_{\bm{k}\lambda}$ is the shorthand for $\theta(\bm{k},\lambda)$.
Considering the relations of Eq.~\eqref{averaging-1}, one can see that only the first three terms of Eq.~\eqref{ret-deltaE-t0} contribute to the interaction energy and the four other terms are vanishing. Hence, we have
\begin{align}
\label{ret-deltaE-t}
\Delta\mathscr{E}_2=-\alpha_2\bigg[
&\Big\langle\bm{\mathcal{E}}\cdot\bm{\mathcal{E}}_{\mu_1}(\bm{R})\Big\rangle
+\frac{1}{2}\Big\langle\bm{\mathcal{E}}_{\mu_1}(\bm{R})\cdot\bm{\mathcal{E}}_{\mu_1}(\bm{R})\Big\rangle
\nonumber\\
+&\Big\langle\bm{E}_0(\bm{R},t)\cdot\bm{E}_{\mu_1}(\bm{R},t)\Big\rangle\bigg]
\ .
\end{align}
The first term of Eq.~\eqref{ret-deltaE-t} corresponds to
the coupling of an electric dipole of the second oscillator
induced by the uniform electric field with the static field of 
the first oscillator, as given by Eq.~\eqref{E-p-static-field}.
The time and phase averaging for this term yields
\begin{align}
\label{Energy-E(p1)-E(xt)}
\Delta\mathscr{E}_2^{(1)}&=-\alpha_2\bm{\mathcal{E}}\cdot\left[\frac{3\hat{z}(\alpha_1\bm{\mathcal{E}}\cdot\hat{z})-\alpha_1\bm{\mathcal{E}}}{[4\pi\epsilon_0]R^3}\right]\nonumber\\
&=\frac{\alpha_1\alpha_2(\mathcal{E}_x^2+\mathcal{E}_y^2-2\mathcal{E}_z^2)}{[4\pi\epsilon_0]R^3}\ .
\end{align}

For the second term of Eq.~\eqref{ret-deltaE-t}, where the static dipole moment of one atom, induced by the electric field of the static dipole moment of the other atom, interacts with the same field, we have
\begin{align}
\label{Energy-E(p1)-E(p1)}
\Delta\mathscr{E}_2^{(2)}&=-\frac{1}{2}\left[\alpha_2\frac{3\hat{z}(\alpha_1\bm{\mathcal{E}}\cdot\hat{z})
	-\alpha_1\bm{\mathcal{E}}}{[4\pi\epsilon_0]R^3}\right]\cdot
\left[\frac{3\hat{z}(\alpha_1\bm{\mathcal{E}}\cdot\hat{z})-\alpha_1\bm{\mathcal{E}}}{[4\pi\epsilon_0]R^3}\right]
\nonumber\\
&=-\frac{\alpha_1^2\alpha_2(\mathcal{E}_x^2+\mathcal{E}_y^2+4\mathcal{E}_z^2)}{2[4\pi\epsilon_0]^2R^6}\ .
\end{align}
The third term of Eq.~\eqref{ret-deltaE-t} describes the interaction energy of two randomly oscillating electric dipoles, which are induced at the corresponding two species by the random zero-point radiation field. Using Eqs.~\eqref{zero-point-field} and \eqref{E-p-rad-field}
for $\bm{E}_0(\bm{R},t)~ \text{and}~\bm{E}_{\mu_1}(\bm{R},t)$, respectively, we obtain \rrr{*ket*}
\begin{widetext}
	\begin{align}
	\label{Energy-E0-E0(p1)-1}
	\Delta\mathscr{E}_3=-\frac{\alpha_1\alpha_2}{[4\pi\epsilon_0]^2}\bigg\langle \sum_{\lambda=1}^2\sum_{\lambda'=1}^2\iint d^3k~d^3k'\ \mathfrak{h}(\bm{k},\lambda) \mathfrak{h}(\bm{k}',\lambda')~ \cos[k'_zR-\Omega' t+\theta(\bm{k}',\lambda')]\qquad\qquad\qquad\qquad\qquad
	\nonumber\\
	\left[\bm{\epsilon}(\bm{k}',\lambda')\cdot\bm{\epsilon}(\bm{k},\lambda)
	\left(
	\frac{k^2}{R}\cos[kR-\Omega\, t+\theta(\bm{k},\lambda)]-
	\frac{k}{R^2}\sin[kR-\Omega\, t+\theta(\bm{k},\lambda)]-
	\frac{1}{R3}\cos[kR-\Omega\, t+\theta(\bm{k},\lambda)]
	\right)
	\right.
	\nonumber\\
	\left.-\epsilon_z(\bm{k}',\lambda')\cdot\epsilon_z(\bm{k},\lambda)
	\left(
	\frac{k^2}{R}\cos[kR-\Omega\, t+\theta(\bm{k},\lambda)]-
	\frac{3k}{R^2}\sin[kR-\Omega\, t+\theta(\bm{k},\lambda)]-
	\frac{3}{R3}\cos[kR-\Omega\, t+\theta(\bm{k},\lambda)]
	\right)
	\right]\rrr{\bigg\rangle}\ ,
	\end{align}
\end{widetext}
which is the same as Eq.~(27) of Ref.~\cite{Boyer1971}.
Taking the same mathematical steps as of Refs.~\cite{Boyer1969} and \cite{Boyer1971}, we arrive at
\begin{equation}
\label{Energy-E0-E0(p1)-2}
\Delta\mathscr{E}_2^{(3)}=-\frac{23\hbar c}{[4\pi\epsilon_0]^2}\frac{\alpha_1\alpha_2}{4\pi R^7}\ ,
\end{equation}
which is the well-known retarded dispersion interaction.
Altogether, this gives us the total interaction energy
\begin{align}
\label{total-E-stoch}
\Delta & \mathscr{E}(R)= \frac{\alpha_1\alpha_2(\mathcal{E}_x^2+\mathcal{E}_y^2-2\mathcal{E}_z^2)}{[4\pi\epsilon_0]R^3} \\ \nonumber
-&\frac{(\alpha_1^2\alpha_2+\alpha_2^2\alpha_1)(\mathcal{E}_x^2+\mathcal{E}_y^2+4\mathcal{E}_z^2)}{2[4\pi\epsilon_0]^2R^6}
-\frac{23\hbar c}{[4\pi\epsilon_0]^2}\frac{\alpha_1\alpha_2}{4\pi R^7}\ ,
\end{align}
where the counterpart of Eq.~\eqref{Energy-E(p1)-E(p1)}, obtained by exchanging $\alpha_1$ and $\alpha_2$, is already added as well.

A remarkable advantage of stochastic electrodynamics is that the origins of all contributions to the total interaction energy can be easily
understood from a (semi)classical point of view. 
Let us interpret our results based on Eqs.~\eqref{ret-deltaE-t} and \eqref{total-E-stoch}. The first term of Eq.~\eqref{total-E-stoch}
is identical to its counterpart in
Eq.~\eqref{nret-total-int-dissimilar} as well as to 
Eqs.~\eqref{ES-pert-mqm} and \eqref{ES-QED} and is the interaction energy of the two static electric dipoles induced by the external field. Therefore, we call it  field-induced electrostatic interaction. 
The second term of Eq.~\eqref{total-E-stoch} is identical to
its counterpart in Eq.~\eqref{nret-total-int-dissimilar} as well as to Eqs.~\eqref{induction-mqm} and \eqref{induction-qed} and corresponds to 
the energy of an oscillator dipole, induced by the electric field of the static field-induced dipole of the other oscillator, interacting with the same field. The nature of this interaction is very similar to the polarization (or induction) interaction between atoms with permanent dipole moments. Hence, we call this term field-induced polarization interaction. 
The third term of Eq.~\eqref{total-E-stoch} describes the well-known Casimir-Polder dispersion interaction between two atoms corresponding to the QED result for the retarded case of large interatomic separations. 
Its nonretarded counterpart is given by the third term of Eq.~\eqref{nret-total-int-dissimilar} and Eq.~\eqref{dispersion-mqm}. 

\begin{figure}[b!]
	\includegraphics[width=0.8\linewidth]{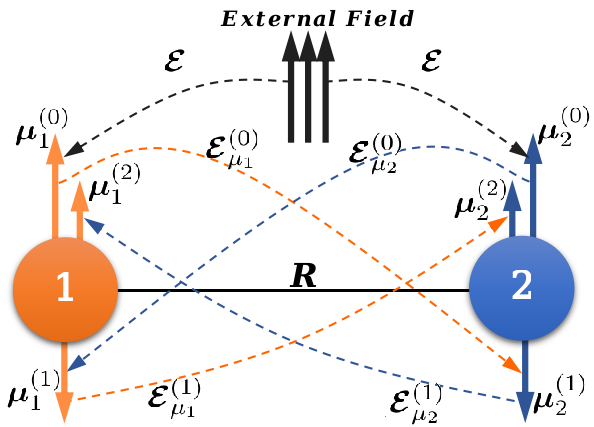}
	\caption{The hierarchy of dipole moments induced by different electric fields are shown for each species: $\bm{\mu_i^{(0)}}$ are the initial dipoles induced by the external field; $\bm{\mu_i^{(1)}}$ are the dipoles induced by the electric fields of the dipoles $\bm{\mu_i^{(0)}}$; $\bm{\mu_i^{(2)}}$ are the dipoles induced by the fields of the dipoles $\bm{\mu_i^{(1)}}$; and so on.}
	\label{fig:induced_dipoles}
\end{figure}

All the above interaction terms are obtained in the lowest order of coupling, as it was assumed in Eq.~\eqref{Energy-in-field}. 
However, it is straightforward to consider higher orders of couplings between the oscillators within the framework of stochastic electrodynamics. 
To this end, one has to take into account the effect of the electric dipoles of the oscillators on the fields that they are interacting with and consider the resulting interactions in a self-consistent approach. 
For instance, when an external static electric field is applied to a system of two interacting atoms or molecules, the external field polarizes them and induces static dipoles, $\bm{\mu_i^{(0)}}$, in each $i$th center of charge, as depicted in Fig.~\ref{fig:induced_dipoles}. 
Thus the zeroth order of coupling, with an interaction energy $U^{(00)}$, occurs between the two externally induced dipole moments, $\bm{\mu_1^{(0)}}$ and $\bm{\mu_2^{(0)}}$. 
In their turn, each of these dipole moments induces another static dipole moment, $\bm{\mu_{i}^{(1)}}$, on the other atom. 
Therefore, the next level of coupling, with an interaction energy $U^{(10)}+U^{(01)}$, occurs between a secondly induced dipole moment of an atom with the dipole moment of the other atom induced by the external field: 
$\bm{\mu_1^{(1)}}\leftrightarrow\bm{\mu_2^{(0)}}$ and $\bm{\mu_1^{(0)}}\leftrightarrow\bm{\mu_2^{(1)}}$. Higher orders of coupling can be described similarly. The total field-induced interaction energy can be obtained from summing up all these different contributions which form an infinite series 
\begin{align}
\label{infinite_series}
U=\sum_{k=0,1,2,...}\ \sum_{l=0,1,2,...} U^{(kl)}\ .
\end{align}
Here, $U^{(kl)}$ denotes the dipole-dipole interaction energy
\begin{align}
U^{(kl)} = \frac{R^2(\bm{\mu}_1^{(k)} \cdot\bm{\mu}_2^{(l)})-3(\bm{\mu}_1^{(k)}\cdot\bm{R})
(\bm{\mu}_2^{(l)}\cdot\bm{R})}{\eta_{k l}\, (4\pi\epsilon_0)\, R^5}\ ,
\end{align}
where $\eta_{k l}$ is a constant prefactor related to the order of induced dipoles. The leading contribution in the series of Eq.~\eqref{infinite_series}, which is $U^{(00)}$ with $\eta_{0 0}=1$, corresponds to the first term of Eq.~\eqref{total-E-stoch}. 
All further contributions to Eq.~\eqref{infinite_series}, with $k>0$ and/or $l>0$, involve dipoles induced by electric fields of other induced dipoles. 
Each time such a dipole moment of one oscillator is induced by an electric field of an induced dipole of another oscillator, where the field is given by Eq.~\eqref{E-p-static-field}. 
The sum of the corresponding first two contributions, $U^{(10)}$ and $U^{(01)}$ with $\eta_{1 0} = \eta_{0 1} = 1/2$, describe the field-induced polarization energy related to the second term of Eq.~\eqref{total-E-stoch}. 
Going further, the sums $U^{(11)}+U^{(20)}+U^{(02)}$ and $U^{(21)}+U^{(12)}+U^{(30)}+U^{(03)}$ correspond, respectively, to the third and fourth terms in Eqs.~\eqref{nret-delta-energy-x-leading} and \eqref{nret-delta-energy-z-leading}. 
This analysis shows that the infinite series of Eq.~\eqref{infinite_series} 
is equivalent to the one obtained in Section III from the exact QM solutions, Eqs.~\eqref{nret-delta-energy-x}--\eqref{nret-delta-energy-z}.

A similar consideration of higher-order couplings between fluctuating dipoles is well-known from literature for the dispersion interactions~\cite{TAD-JCP2013}.
The lowest order of coupling occurs between fluctuating dipole moments induced by the random zero-point radiation field and results in the London/Casimir-Polder dispersion interaction for the nonretarded/retarded regime. 
Due to the employment of the QDO model, our exact diagonalization approach successfully captures all such higher-order coupling terms on equal footing for both, the field-induced electrostatic/polarization and dispersion interactions.

\section{Application to atomic and molecular systems}
In this section, we apply the derived formulas to nucleo-electronic systems, considering argon-argon and benzene-benzene as two representative examples for atomic and molecular dimers, respectively. 
The chosen systems allow us to study field-induced effects on intermolecular interactions in systems of varying polarizability, for different configurations of the considered dimers and the applied electric field. With these examples, we illustrate the possibility to switch between molecular conformations and dissociate molecular dimers with an external electric field.
In what follows, we discuss the three contributions to the total force, $F = - \nabla_R [\Delta\mathscr{E}(R)]$,  stemming from field-induced electrostatics, field-induced polarization, and dispersion contributions to the interaction energy, $\Delta\mathscr{E}(R)$. 
The latter is given by  Eqs.~\eqref{nret-total-int-dissimilar} and \eqref{total-E-stoch} for the nonretarded and retarded cases, respectively. 
The obtained either negative or positive forces correspond to the attractive and repulsive interactions, respectively. 
Among the three forces, the field-induced polarization and dispersion forces always remain attractive, whereas the field-induced electrostatic force can change its sign depending on the direction of the applied electric field with respect to the line connecting the two species. 
This force is attractive when the field is applied along the inter-species separation, and is repulsive when the field is perpendicularly applied to the dimer. 
The obtained three forces scale as $\propto \alpha^2 \mathcal{E}^2/R^4$, $\propto \alpha^3 \mathcal{E}^2/R^7$, and $\propto \alpha^2 \hbar\omega/R^7$ ($\propto \alpha^2 \hbar c/R^8$) for the field-induced electrostatic, field-induced polarization, and nonretarded (retarded) dispersion interactions, respectively. 
From these scaling laws, it follows that the field-induced polarization force can become comparable to the field-induced electrostatic force only for systems with high polarizabilities. 
In addition, the two field-induced forces similarly depend on the strength of the applied electric field, whereas the dispersion force does not depend on it.

\begin{figure}[t]
	\includegraphics[width=\linewidth]{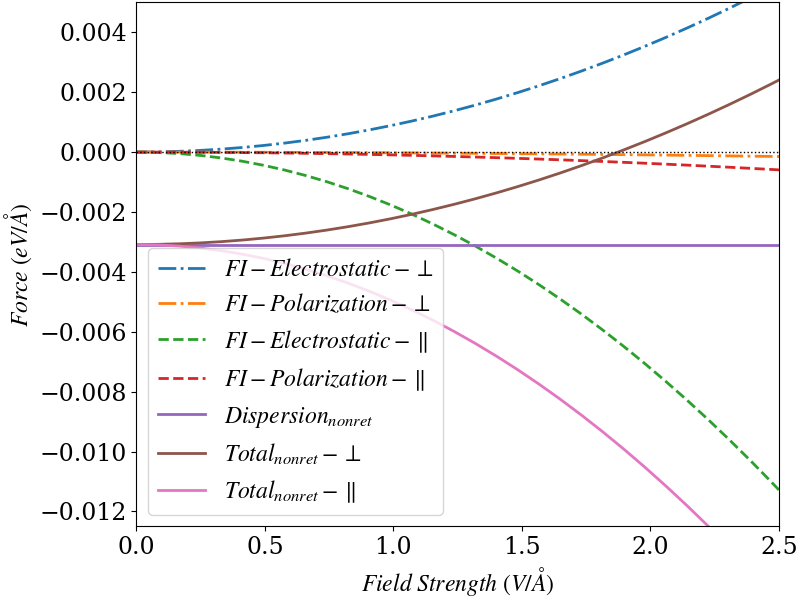}
	\caption{Nonretarded interatomic forces for two interacting argon atoms separated by $R$=5\,\AA. 
	The symbols $||$ and $\perp$ indicate the two cases when the field is either parallel or perpendicular to the line connecting the centers of the atoms. For a field of the strength $\approx$ 1.8\,V/\AA\  perpendicularly applied to the dimer, the repulsive field-induced (FI) electrostatic force compensate the attractive field-induced polarization and dispersion forces.}
	\label{fig:Argon_vdW_nonret}
\end{figure}

First, we consider the argon dimer. The atomic dipole polarizability of argon, $\alpha = 11.1$~a.u.~\cite{Jones2013}, is quite small. Consequently, the field-induced forces (especially, the field-induced polarization force) are weak for this system. In order to obtain reasonable force values, we restrict our consideration to the nonretarded case corresponding to smaller interatomic separations. 
Since, for argon, $\omega_e = 0.7272$~a.u.~\cite{Jones2013}, one has 
$\lambda_e = 2\pi c/\omega_e = 1183.7$~a.u.~$\approx 626$\,\AA. For our analysis, we have chosen the interatomic distance $R$ = 5\,\AA, which corresponds to the nonretarded regime, $R \ll 600$\,\AA. 
Figure~\ref{fig:Argon_vdW_nonret} shows that, for two argon atoms separated by the chosen distance, the field-induced polarization force becomes negligible in comparison to the dispersion and field-induced electrostatic forces.

\begin{figure}[b!]
\includegraphics[width=0.9\linewidth]{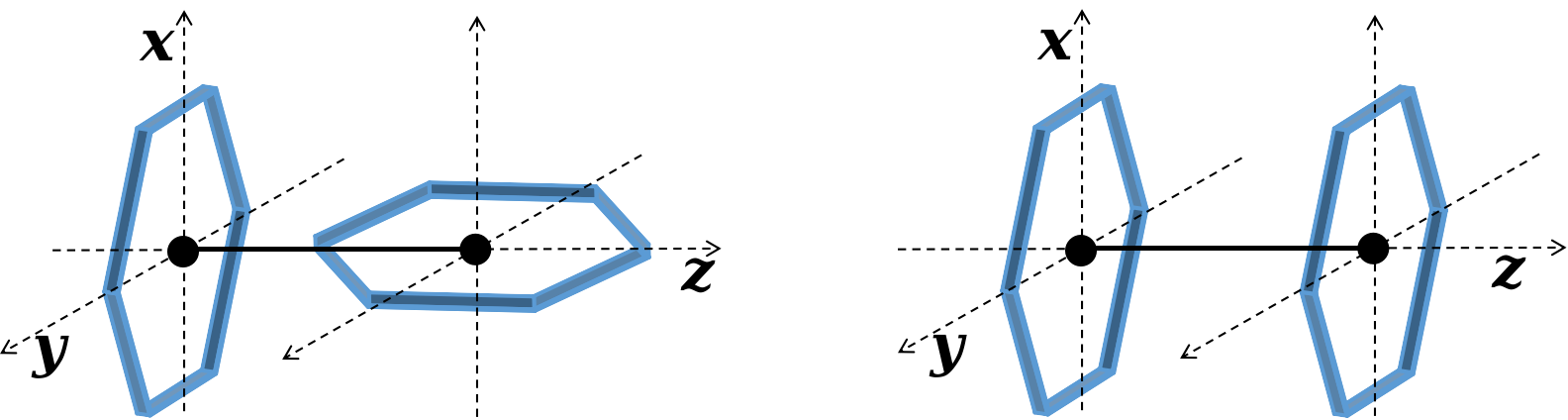}
	\caption{Two configurations of benzene dimers; {\bf left}: T-shaped structure $T (C_{2v})$, and {\bf right}: Sandwich structure $SW (D_{6h})$.}
	\label{fig:Benzene_str}
\end{figure}

\begin{figure}[t]
	\includegraphics[width=\linewidth]{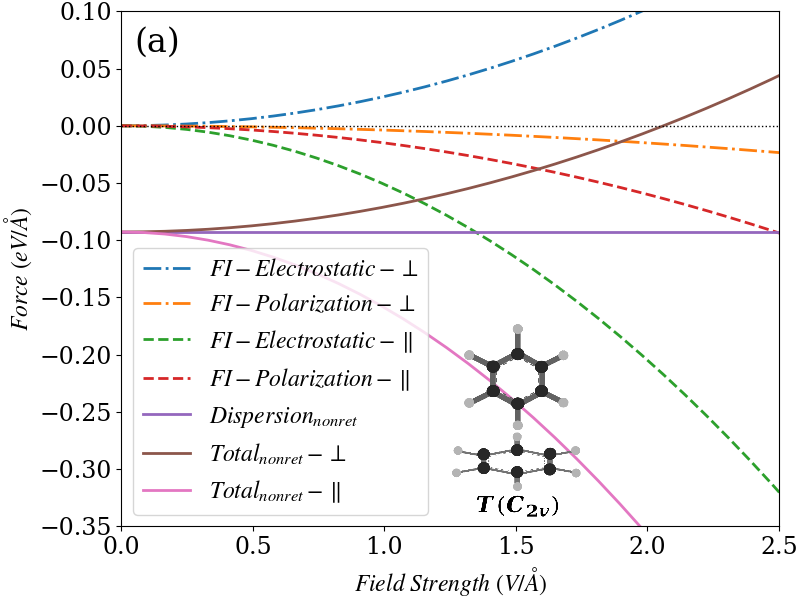}
	\\ \vspace{0.25cm}
	\includegraphics[width=\linewidth]{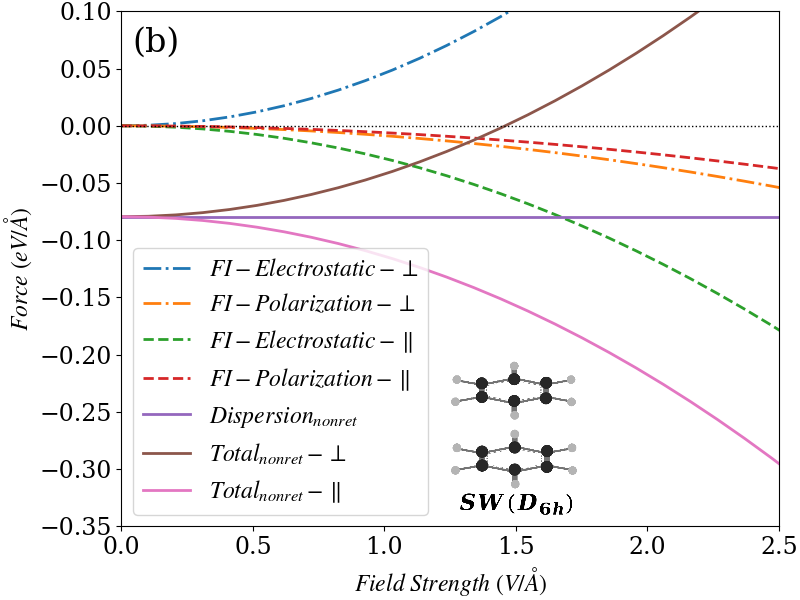}
	\caption{Nonretarded intermolecular forces for two interacting benzene molecules separated by a distance of $R=5\,$\AA\ and possessing {\bf (a)} T-Shape or {\bf (b)} Sandwich structure. 
	The symbols $||$ and $\perp$ indicate the field applied
	either parallel or perpendicular to the line connecting the centers of the molecules. For an external field of the strength $\mathcal{E}\approx$ 2\,V/\AA\  perpendicularly applied to a T-shaped benzene dimer the repulsive field-induced (FI) electrostatic force compensate the attractive field-induced polarization and dispersion forces while such compensation in Sandwich structure of benzene dimer occurs at $\mathcal{E}\approx$ 1.5\,V/\AA .}
	\label{fig:Benzene_vdW_nonret}
\end{figure}

\begin{figure}[t]
	\includegraphics[width=\linewidth]{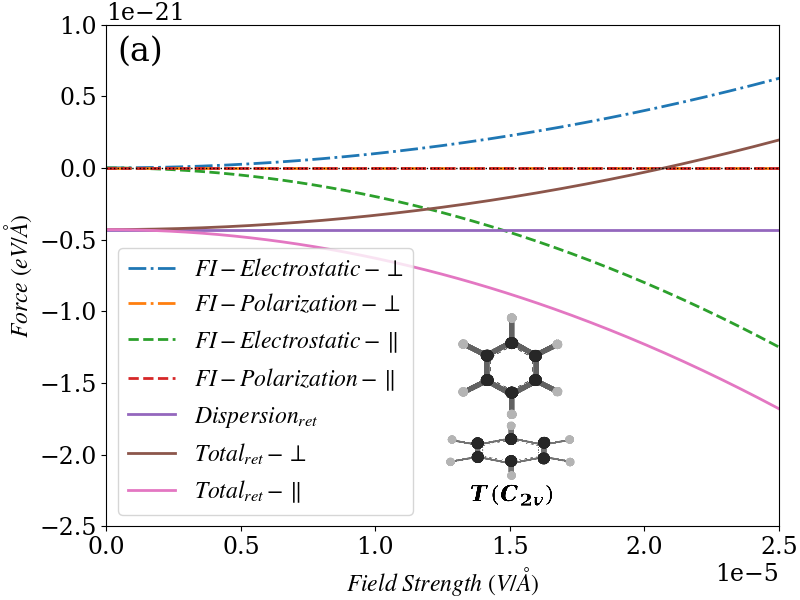}
	\\ \vspace{0.25cm}
	\includegraphics[width=\linewidth]{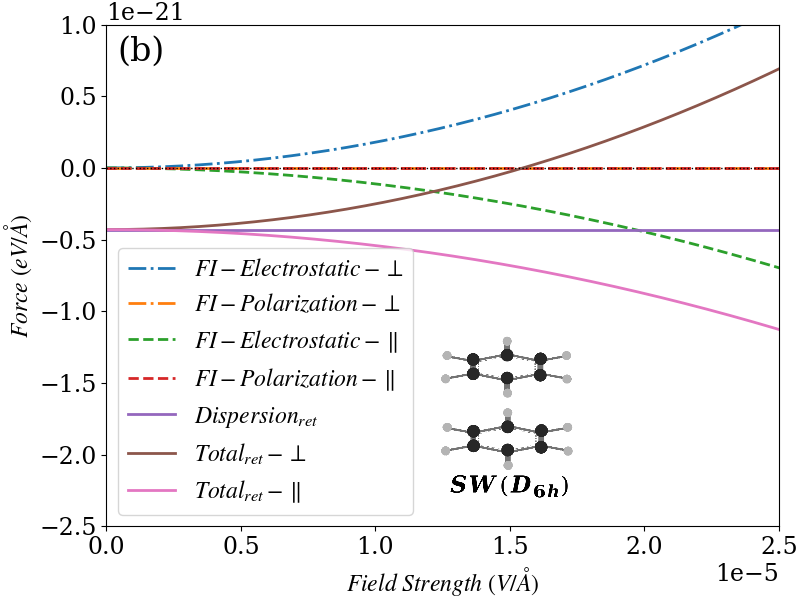}
	\caption{Retarded intermolecular forces for two interacting benzene molecules separated by a distance of $R=2000\,$\AA\ and possessing {\bf (a)} T-Shape or {\bf (b)} Sandwich structure. 
	The symbols $||$ and $\perp$ indicate the field applied
	either parallel or perpendicular to the line connecting the centers of the molecules. For an external field of the strength $\mathcal{E}\approx 2\times10^{-5}\,$V/\AA\ perpendicularly applied to a T-shaped benzene dimer, the repulsive field-induced (FI) electrostatic force compensate the attractive field-induced polarization and dispersion forces while such compensation in Sandwich structure of benzene dimer occurs at $\mathcal{E}\approx 1.5\times10^{-5}\,$V/\AA .}
	\label{fig:Benzene_vdW_ret}
\end{figure}

Therefore, it is enough to take into account the latter two forces only. 
Their strength is governed by an interplay between how large is the interatomic distance and how strong is the static electric field. 
In addition, the external field can be applied in two qualitatively different directions, parallel and perpendicular to the line connecting the two argon atoms. 
For the static electric field applied along the interatomic distance, the field-induced electrostatic interaction is attractive and it can only enhance the dispersion attraction. 
By contrast, for the electric field applied perpendicular to the interatomic distance, the field-induced electrostatics becomes repulsive which makes it competitive with the dispersion attraction. 
For this case, at the field strength of about 1.8\,V/\AA, the net force vanishes.

Let us now consider the interaction of two molecules. 
The benzene dimers have been often used as one of the simplest systems to study vdW interactions involving two aromatic molecules of $\pi\!-\!\pi$ type, which play a key role in chemistry and biology. Here, we apply a uniform static electric field to two different configurations of the benzene dimer, namely the T-shaped structure with $C_{2v}$ symmetry ($T (C_{2v})$) and the Sandwich structure with $D_{6h}$ symmetry ($SW (D_{6h})$), as illustrated in Fig.~\ref{fig:Benzene_str}.

The in-plane, out-of-plane, and average (isotropic) dipole polarizabilities of a benzene molecule are well-known~\cite{DiStasio2014} as (in atomic units) $\alpha_{\rm in}=82.00$, $\alpha_{\rm out}=45.10$, and $\alpha_{\rm avg}=\frac{1}{3}(\alpha_{xx}+\alpha_{yy}+\alpha_{zz})=69.70$,  respectively. Then, the QDO characteristic frequency of benzene can be computed according to Eq.~\eqref{QDO-parameter} as
\begin{equation*}
\omega_e = {4\, C_6}/{(3\, \hbar\, \alpha_{\rm avg}^2)} = 0.4729~{\rm a.u.}\ ,
\end{equation*}
where the dispersion coefficient of the benzene-benzene vdW interaction, $C_6 = 1723$~a.u., is taken from Ref.~\cite{Meath1992}. Similarly, the corresponding wavelength is obtained as
\begin{equation*}
\lambda_e = {2\pi c}/{\omega_e} =
1820~{\rm a.u.} \approx 963\, \text{\AA}\ .
\end{equation*}
Consequently, for the intermolecular distances in the benzene dimer such that $R\ll 10^3$\AA\ or $R\gg 10^3$\AA\ we have the nonretarded or retarded interactions, respectively.

For a varying strength of a uniform static electric field applied to the benzene dimers, in Figs.~\ref{fig:Benzene_vdW_nonret} and \ref{fig:Benzene_vdW_ret} we show
the intermolecular forces for the nonretarded ($R=5\,$\AA) and retarded ($R=2000\,$\AA) regimes, respectively. 
The magnitude of the intermolecular forces (as well as of the strength of the applied static electric fields) in the retarded regime is drastically smaller compared to the nonretarded regime. Nevertheless, by comparing Figs.~\ref{fig:Benzene_vdW_nonret} and \ref{fig:Benzene_vdW_ret}, one can see the same qualitative behavior for both regimes. Although being negligible for the considered benzene dimers in practice (see Fig.~\ref{fig:Benzene_vdW_ret}), the intermolecular forces corresponding to the retarded regime can become measurable for the case of extended (bio)molecules possessing large polarizabilities. 
The presented results show that in both, nonretarded and retarded cases, the total field-induced force can overtake the dispersion force for certain strengths of the static field, if the latter is applied perpendicular to the intermolecular distance. 
The field strength at which the field-induced forces and the dispersion force cancel out depends on the intermolecular distance and the structure of the dimer. 
At any separation distance, the needed field strength for such a compensation is always smaller for the SW structure in comparison to the T-shaped structure.

\begin{figure}[t!]
		\includegraphics[width=\linewidth]{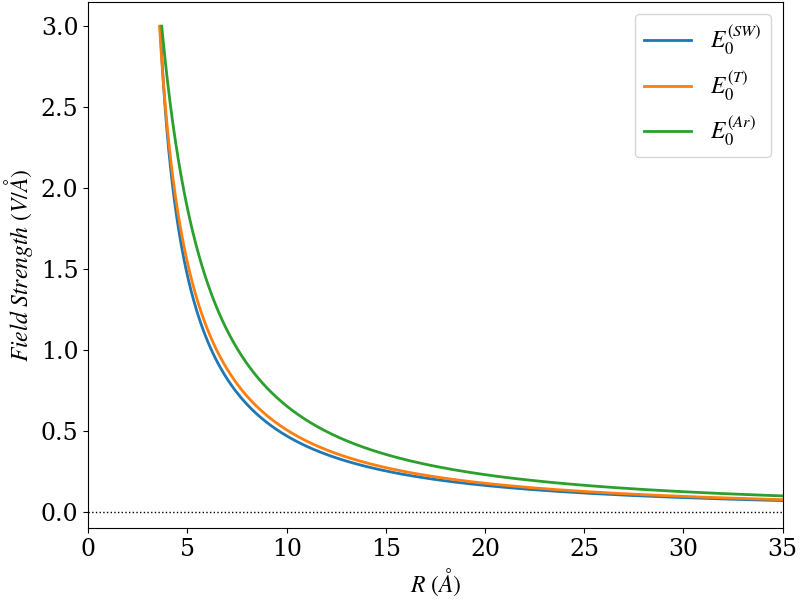}
	\caption{Strength of a static electric field, perpendicularly applied to benzene and argon dimers, at which the field-induced and dispersion forces cancel out, is shown versus intermolecular/interatomic distance (represented in logarithmic scale). The nonretarded regime of the molecular interactions is considered, which corresponds to the results of Figs.~\ref{fig:Argon_vdW_nonret} and \ref{fig:Benzene_vdW_nonret}.}
	\label{fig:Strength_Ar_benzene}
\end{figure}

Figure~\ref{fig:Strength_Ar_benzene} shows the strength of
the static field at which the net force vanishes versus interspecies distance, for the case when the field is perpendicularly applied to the benzene and argon dimers. 
As $R$ increases (starting from values close to equilibrium distances in the dimers) the strength of the compensating field ($E_0$) becomes smaller. 
For a range of $R$ which is more probable in stable dimers (slightly larger than the equilibrium distance of the dimers in the absence of the external field), $E_0$ is always larger for the Ar-Ar system compared to both, T--shaped and Sandwich structures of benzene dimer. 
This difference indicates that electric fields from external sources and nearby molecules should have a stronger influence on larger molecules.

\section{Discussion and Conclusion}
We have employed the QDO model, as an efficient tool for describing atomic/molecular polarization response, to derive different forces acting on two separated species (atoms or molecules) under the combined action of a static--electric and vacuum--radiation fields. 
The obtained three lowest-order (with respect to the inverse separation distance, $R^{-1}$) contributions to these forces stem from the field-induced electrostatics ($\propto R^{-3}$), field-induced polarization ($\propto R^{-6}$), and dispersion ($\propto R^{-6}/R^{-7}$) interactions. 
All the three contributions to the interaction energy form an infinite series due to the self-consistent mutual polarization of the interacting species (see Sections III and VI).
The field-induced interactions are not influenced by the retardation effects, whereas the dispersion interaction shows a conventional behavior for nonretarded and retarded regimes, both of which are not affected by static electric fields. 
For the considered unconfined atoms in isotropic and homogeneous vacuum, the field-induced polarization and dispersion forces remain attractive. In contrast, the field-induced electrostatic force becomes attractive or repulsive for the electric field applied either along the separation distance or perpendicular to it, respectively. 
Therefore, it is possible to tune the intermolecular interactions by a variation of the strength and the direction of the applied electric field. 
In order to resolve many existing discrepancies and strengthen partial results available in the literature, our comprehensive framework is based on four complementary approaches rooted in quantum mechanics, quantum electrodynamics, and stochastic electrodynamics. 
The employment of these four approaches leads to a systematic and robust characterization of intermolecular interactions under the combined action of an externally-applied field and the ever-present vacuum field. 
A generalization of the presented framework to many QDOs, higher multipole contributions, as well as to the case of spatially-confined systems can be performed in a straightforward manner.

In order to assess the potential of our framework for practical applications, we have considered and compared argon-argon and benzene-benzene dimers, as representatives models for atomic and molecular systems. We showed that the field-induced polarization plays a minor role for the considered dimers. 
However, the polarization contribution can become important for highly-polarizable systems (especially, systems excited by optical modes with frequencies close to the molecular characteristic frequencies) since field-induced polarization scales with the cube of the dipole polarizability, whereas the other two forces scale with the square of the dipole polarizability.
Generally, the effect of a static electric field can be assumed negligible for small atomic systems since the field-induced electrostatic force can compete with the dispersion one only at large separations for reasonable electric fields much weaker than the internal atomic one. 
However, the situation becomes more intricate for large molecular systems, especially at the nanoscale. 
Here, the effective normal-mode polarizabilities become highly anisotropic and can easily reach 2-3 orders of magnitude higher values than those of small molecules~\cite{Ambrosetti-Science}. 
This may lead to a non-trivial interplay between field-induced forces with the dispersion one. 
In addition, when increasing the size of the system, the effective separation distance between its components becomes larger. 
In turn, the increased separation enhances the field-induced electrostatic force with respect to the other forces, which can either amplify or weaken the intermolecular interactions depending on the direction of an applied field. 
Consequently, we suggest that the action of external electric fields should become relevant for macromolecules and nanoscale objects.

It is important to embed our derivations and results into the state of the art in the literature. 
As was mentioned above, for the field-induced forces to become comparable to the dispersion force at short separation distances, one needs to apply quite strong electric fields in case of atomic systems. 
Indeed, the effective electric field acting from the argon nucleus on its valence shell is $\propto$ 10~V/{\AA}. 
On the other hand, the field-induced electrostatic force in the Ar-Ar dimer with $R=5$~{\AA} becomes comparable to the dispersion force at an external field of $\propto 1$~V/{\AA}. 
Thus, for reasonable strengths of external electric fields, the field-induced forces are not relevant in the case of vdW-bonded atomic systems. 
This statement is in agreement with the conclusion of Ref.~\cite{Milonni1996}, where the leading contribution to the field-induced electrostatic interaction was derived based on classical electromagnetic theory. 
Nevertheless, the field strength required to make the field-induced electrostatic force comparable to the dispersion one rapidly decreases with increasing $R$, as illustrated by Fig.~\ref{fig:Strength_Ar_benzene}. 
Consequently, at large interatomic distances, particularly in the retarded regime, the field-induced interactions can become dominant even for weak applied fields. 
In addition, increased field effects are expected for large molecular systems. 
As was already mentioned above, in such systems the many-body effects can drastically influence the strength of the interaction and therefore much weaker applied fields can cause strong effects. 
A specifically interesting case is when an external electric field is due to a single optical mode. 
As discussed in Ref.~\cite{Milonni1996}, the  difference present for the field-induced electrostatic interaction in that case can be effectively described by replacing the static polarizability $\alpha \equiv \alpha(0)$ by its frequency-dependent counterpart $\alpha (\omega_{\rm opt})$, where $\omega_{\rm opt}$ is the frequency of the optical mode. 
Thus, by choosing a proper optical frequency, one can drastically enhance the polarizability $\alpha (\omega_{\rm opt}) \propto (\omega - \omega_{\rm opt})^{-1}$. 
Taking this into account, we expect that for large molecules such a setup can significantly increase the role of the field-induced polarization force, as not considered in Ref.~\cite{Milonni1996} but revealed within our work.

The fact that, within the considered lowest order of coupling between matter and vacuum radiation field, the dispersion interactions between two atoms or molecules are not affected by a static electric field stems from their quantum-mechanical nature. 
Since an applied uniform static field influences fluctuations of neither the vacuum radiation field nor electronic densities, it cannot affect the considered dispersion interactions.
This implies that, neglecting higher-order contributions, like from field-induced hyperpolarizabilities of atoms scaled with $R^{-11}$~\cite{Salam1997,Hu2021}, under static electric fields, the leading contributions to the dispersion energy remain $\propto R^{-6}$ and $\propto R^{-7}$ for the nonretarded and retarded regimes, respectively. 
Within the perturbative technique of the QED theory, these dispersion interactions arise from the 4th-order of the coupling of matter to the vacuum radiation field and the two interacting atoms exchange a pair of virtual photons. 
However, the above commonly accepted picture was recently questioned by Fiscelli~\emph{et~al.}~\cite{Fiscelli2020} who obtained, within the 2nd-order of perturbation, the dispersion interaction energy between two atoms under static electric fields as $\propto R^{-3}$ and $\propto R^{-4}$ for nonretarded and retarded regimes, respectively. 
A careful consideration of the approach used by Fiscelli~\emph{et~al.}~\cite{Fiscelli2020} can identify an error in their analysis caused by employing perturbation theory in two steps. 
Namely, in Ref.~\cite{Fiscelli2020}, first the wavefunctions of a two-level ``hydrogen'' atom in a static electric field were obtained from perturbation theory, by considering the external field as a perturbation. 
Then, the obtained wavefunctions were used as unperturbed eigenstates of an atom (under the static field) to be coupled to another ``hydrogen" atom through the vacuum radiation field. 
Considering this coupling as a new perturbation, Fiscelli~\emph{et~al.}~\cite{Fiscelli2020} used perturbation theory for the second time. 
However, as mentioned above, the ``unperturbed'' wavefunctions employed for this step, were obtained in Ref.~\cite{Fiscelli2020} from the first use of perturbation theory by the authors. 
As a result, these ``unperturbed" wavefunctions do not form a complete set and, strictly speaking, they cannot be used for expanding the eigenstates of the system of two interacting ``hydrogen" atoms under a static electric field. 
This incompletness of the wavefunctions seems to be the origin of the unusual scaling law of $R^{-4}$ obtained in Ref.~\cite{Fiscelli2020} for the retarded regime, as was also suggested in Ref.~\cite{Hu2021} published after the initial submission of our current manuscript. Indeed, we have found
that, by applying the Gram-Schmidt orthonormalization procedure to ``hydrogen" wavefunctions under a static electric field obtained by Fiscelli~\emph{et~al.}~\cite{Fiscelli2020}, their term $\propto R^{-4}$ transforms to $\propto R^{-3}$. 
Hence, there should be no influence of the retardation on the interaction energy obtained in Ref.~\cite{Fiscelli2020}, which already suggests that the $R^{-3}$ term derived in that work is of electrostatic origin. 
In addition, we emphasize the fact that the interaction energy was obtained in Ref.~\cite{Fiscelli2020} from the 2nd-order of the QED perturbation theory. 
Taking into account our detailed derivation performed within Section V, one can finally conclude that the (corrected) results of Fiscelli~\emph{et~al.}~\cite{Fiscelli2020} correspond to our field-induced electrostatic interaction.

The above discussion underlines the importance of robust and comprehensive frameworks such as the one developed within the presented work. Based on molecular quantum mechanics and quantum electrodynamics, our framework employs the QDO model as a well-established coarse-grained formalism to describe electronic response properties and dispersion interactions. 
Unlike the ``two-level atom'' model, widely used in quantum optics and quantum electrodynamics, the QDO model allows exact solutions under the effects of a variety of external fields and/or boundary conditions. With the developed extension of this efficient model to the presence of external static electric fields, our framework paves the way for a deeper understanding of inter- and intra-molecular interactions under various electromagnetic fields. The further possible studies can capture  considerations of nontrivial effects of geometric confinements and boundary conditions on these interactions, with an eventual practical use of such knowledge especially in chemistry, nanoscience, and biophysics.
The derived formalism provides a reliable picture of the field-induced and dispersion interactions going from the nonretarded to retarded regime, and is amenable to various extensions from the two-body to many-body interactions between atoms or molecules. Indeed, the analytical solution given by  Eq.~\eqref{nret-total-int-dissimilar} can be straightforwardly generalized to any number of QDOs, each of them under a different static field. The latter approach would allow to effectively model internal atom-dependent electric fields present in large molecules.

As a brief summary, we enumerate several potential implications and possible extensions of our work:
\begin{itemize}
\item \textit{Employing the QDO model within QM and QED theory of intermolecular interactions enables studying atomic and molecular systems under the influence of external sources or fields.}\\
Due to the quadratic form of the QDO Hamiltonian, the problem of coupling this quantum-mechanical system to external fields and/or boundary conditions is analytically solvable within the dipole approximation or the multipole expansion of the Coulomb potential. Hence, one can perform perturbative QM and QED calculations of intermolecular interactions between atoms or molecules. This allows to investigate retarded and nonretarded interactions in molecular systems of increasing complexity and unambiguously classify the different types of field-induced molecular interactions.

\item \textit{In the nonretarded regime, the effect of external fields on intermolecular interactions can be straightforwardly generalized to systems with an arbitrary number of interacting species by implementing field-induced changes in a system of many interacting QDOs.}\\
As discussed above, an arbitrary number of QDOs coupled through the dipole-dipole potential under a static electric field is an exactly solvable problem in quantum mechanics.
Taking into account the field-induced redistributions of electron densities in many-body systems, one can investigate the effect of external fields on many-body interactions.

\item \textit{Using the QDO model, one can capture the effect of intramolecular local fields in large molecules.}\\
The opportunity to diagonalize the total Hamiltonian of QDOs under a static field, like in Eq.~\eqref{nret-Hamiltonian-x4}, implies that using the QDO model one can also capture the effect of intramolecular fields acting on atoms in a molecule. Indeed, covalent interactions cause charge transfers between atoms, which leads to a redistribution of local centers of positive and negative charges over the molecular space. 
This effect can be described via local effective external fields acting on atoms. Using our exact-diagonalization method, one can take into account the effect of such fields via spatial shifts of the QDO centers of oscillation. Such generalization would extend our framework to the study of intramolecular interactions.

\item \textit{The dispersion interaction between two atoms, as a result of quantum-mechanical fluctuations of the electronic density, is not affected by external uniform static fields.}\\
Dispersion interactions originate from quantum-mechanical fluctuations of electronic structures of matter and the vacuum field. Consequently, these interactions cannot be influenced by uniform static fields. However, static fields inducing electrostatic and polarization interactions, can qualitatively and quantitatively change total intermolecular interactions. These hypotheses were comprehensively investigated and confirmed in the present work by employing four complementary approaches. 

\item \textit{Employing the QDO model for electronic polarization response allows one to better understand and classify QED effects in atoms and molecules.}\\
Perturbation theory, as a powerful mathematical tool, is widely used in QM and QED, including its various applications in physics and chemistry. This approach considers the effects of small perturbations on the properties of a QM system. Within quantum mechanics, this implies that states of the perturbed system can be expanded in terms of states of the unperturbed system, which requires the latter to form a complete set. Employing the QDO model, as an exactly solvable problem under a variety of physical conditions, enables us to apply straightforward perturbation theory techniques to coupled QDOs and obtain robust classification of different types of field-induced molecular interactions. This is an especially interesting approach to search for non-trivial field-induced interactions in QED and quantum-field theory.\\

\item \textit{Intermolecular interactions can be tailored by applying static electric fields, which induce field-dependent electrostatic and polarization forces.}\\
Attractive/repulsive character of the obtained field-induced electrostatic force depends on the orientation of the applied field with respect to the separation distance while the field-induced polarization force is always attractive. When the external field is applied perpendicularly, the field-induced electrostatic force becomes repulsive. In such case, the interplay between the field--induced and dispersion forces can be used as a mechanism for controlling intermolecular interactions.
\end{itemize}

In summary, we derived and discussed four complementary formalisms, which constitute a robust framework for investigating molecular interactions at arbitrary separation distances under the influence of uniform static electric fields. We showed that such fields induce static atomic polarization, offering an opportunity to tune molecular interactions via an interplay of field-induced electrostatics/polarization as well as dispersion interactions. To conclude, we remark that our framework barely scratches the surface of possible developments and applications in the field of molecular interactions under the combined action of external and vacuum fields.

\section*{ACKNOWLEDGMENTS}
The authors acknowledge the financial support from the Luxembourg National Research Fund through the FNR CORE projects ``QUANTION(C16/MS/11360857, GrNum:11360857)" and
``PINTA(C17/MS/11686718)" as well as from the European Research Council via ERC Consolidator Grant ``BeStMo(GA n725291)".

\begin{widetext}
\appendix
\section{The case of anisotropic QDOs}
In order to extend our result for the interaction energy given by Eq.~\eqref{nret-total-int-dissimilar} to the case of anisotropic molecules, like benzene, one has to take into account an anisotropy of the polarizability. This quantity plays the role of a coupling constant of an atom or molecule to an electric field. 
Generally, the dipole polarizability is a second-rank tensor, which can be diagonalized using the principal axes. 
By choosing the Cartesian coordinate system along such axes, we obtain
\begin{align}
\label{A-1}
\Delta \mathscr{E} =
&\frac{1}{[4\pi \epsilon_0] R^3}
\bigg\{
\alpha_{xx}^{(1)} \alpha_{xx}^{(2)}~\mathcal{E}_x^2+
\alpha_{yy}^{(1)} \alpha_{yy}^{(2)}~\mathcal{E}_y^2-
2\alpha_{zz}^{(1)} \alpha_{zz}^{(2)}~\mathcal{E}_z^2
\bigg\}
\nonumber\\
-&\frac{1}{2 [4\pi\epsilon_0]^2 R^6}
\bigg\{
\alpha_{xx}^{(1)} \alpha_{xx}^{(2)}[\alpha_{xx}^{(1)} +\alpha_{xx}^{(2)}] ~\mathcal{E}_x^2+
\alpha_{yy}^{(1)} \alpha_{yy}^{(2)}[\alpha_{yy}^{(1)} +\alpha_{yy}^{(2)}] ~\mathcal{E}_y^2+
4\alpha_{zz}^{(1)} \alpha_{zz}^{(2)}[\alpha_{zz}^{(1)} +\alpha_{zz}^{(2)}] ~\mathcal{E}_z^2
\bigg\}
\nonumber\\
-&\frac{\hbar}{4 [4\pi\epsilon_0]^2 R^6}
\left(\frac{\omega_1\omega_2}{\omega_1+\omega_2}\right)
\bigg\{
\alpha_{xx}^{(1)} \alpha_{xx}^{(2)}+
\alpha_{yy}^{(1)} \alpha_{yy}^{(2)}+
4\alpha_{zz}^{(1)} \alpha_{zz}^{(2)}
\bigg\}
\ ,
\end{align}
for the nonretarded interaction energy between two molecules. Here, $\alpha_{ii}^{(n)}$ denotes $ii$th Cartesian component of the polarizability tensor of the $n$th molecule. Equation \eqref{A-1} {as well as its retarded counterpart, straightforwardly obtained by a similar generalization of Eq.~\eqref{total-E-stoch}, were used} in Section VII to compute dispersion forces for the benzene dimers.

\section{Dissimilar local static electric fields applied to isotropic QDOs}
If the two interacting QDOs undergo locally different static fields, $\bm{\mathcal{E}}_1 = (\mathcal{E}_{1x},\mathcal{E}_{1y},\mathcal{E}_{1z})$ and $\bm{\mathcal{E}}_2 = (\mathcal{E}_{2x},\mathcal{E}_{2y},\mathcal{E}_{2z})$, the field--induced contributions to the interaction energy of Eqs.~\eqref{nret-total-int-dissimilar}, \eqref{ES-QED}, \eqref{induction-qed}, and \eqref{total-E-stoch} take the following forms 
\begin{align}
\label{DeltaE-4-3-1_iso}
\Delta \mathscr{E}_{\rm FI} = & \Delta \mathscr{E}_{\rm FI}^{\rm el} \left(R^{-3}\right) + \Delta \mathscr{E}_{\rm FI}^{\rm pol} \left(R^{-6}\right) =
\frac{1}{[4\pi\epsilon_0] R^3}\bigg\{
(\alpha_{1}\mathcal{E}_{1x})\,
(\alpha_{2}\mathcal{E}_{2x})+
(\alpha_{1}\mathcal{E}_{1y})\,
(\alpha_{2}\mathcal{E}_{2y})-
2\,(\alpha_{1}\mathcal{E}_{1z})\,
(\alpha_{2}\mathcal{E}_{2z})
\bigg\} \\ \nonumber
-&\frac{1}{2\,[4\pi\epsilon_0]^2 R^6}\bigg\{\alpha_2\Big[
(\alpha_{1}\mathcal{E}_{1x})^2 +
(\alpha_{1}\mathcal{E}_{1y})^2 +
4\,(\alpha_{1}\mathcal{E}_{1z})^2 \Big] + \alpha_1\Big[
(\alpha_{2}\mathcal{E}_{2x})^2 +
(\alpha_{2}\mathcal{E}_{2y})^2 +
4\,(\alpha_{2}\mathcal{E}_{2z})^2 \Big] \bigg\}\ ,
\end{align}
where, for simplicity, we assume both QDOs to be isotropic: $\alpha_1 = \alpha_{xx}^{(1)} = \alpha_{yy}^{(1)} = \alpha_{zz}^{(1)}$ and $\alpha_2 = \alpha_{xx}^{(2)} = \alpha_{yy}^{(2)} = \alpha_{zz}^{(2)}$. This setup is similar to the one of Ref.~\cite{Fiscelli2020}, where locally different static fields but isotropic polarizabilities were used.

\section{Dissimilar local static electric fields applied to anisotropic QDOs}
Finally, the most general case describes two interacting anisotropic QDOs undergoing locally different static fields. For this situation, the corresponding field-induced interactions present in Eqs.~\eqref{nret-total-int-dissimilar}, \eqref{ES-QED}, \eqref{induction-qed}, and \eqref{total-E-stoch} transform to 
\begin{align}
\label{DeltaE-4-3-1_aniso}
\Delta \mathscr{E}_{\rm FI} = &
\frac{1}{[4\pi\epsilon_0] R^3}\bigg\{
(\alpha_{xx}^{(1)}\mathcal{E}_{1x})\,
(\alpha_{xx}^{(2)}\mathcal{E}_{2x})+
(\alpha_{yy}^{(1)}\mathcal{E}_{1y})\,
(\alpha_{yy}^{(2)}\mathcal{E}_{2y})-
2\,(\alpha_{zz}^{(1)}\mathcal{E}_{1z})\,
(\alpha_{zz}^{(2)}\mathcal{E}_{2z})
\bigg\}\nonumber\\
-&\frac{1}{2\,[4\pi\epsilon_0]^2 R^6}\bigg\{\Big[
\alpha_{xx}^{(2)}\,(\alpha_{xx}^{(1)}\mathcal{E}_{1x})^2 +
\alpha_{yy}^{(2)}\,(\alpha_{yy}^{(1)}\mathcal{E}_{1y})^2 +
4\,\alpha_{zz}^{(2)}\,(\alpha_{zz}^{(1)}\mathcal{E}_{1z})^2
\Big]\nonumber\\
&\qquad\qquad\quad\!
+\Big[
\alpha_{xx}^{(1)}\,(\alpha_{xx}^{(2)}\mathcal{E}_{2x})^2 +
\alpha_{yy}^{(1)}\,(\alpha_{yy}^{(2)}\mathcal{E}_{2y})^2 +
4\,\alpha_{zz}^{(1)}\,(\alpha_{zz}^{(2)}\mathcal{E}_{2z})^2 \Big]
\bigg\}\ ,
\end{align}
which can be simply obtained as a combination of Eqs.~\eqref{A-1} and \eqref{DeltaE-4-3-1_iso}. Equation~\eqref{DeltaE-4-3-1_aniso}, together with the dispersion contribution of Eq.~\eqref{A-1} or its retarded counterpart, provides one with a practical tool to study intermolecular interactions in various nucleo-electronic systems mentioned in Section VIII.
\\
\end{widetext}


\end{document}